\documentclass[12pt,reqno]{amsart}

\usepackage[applemac]{inputenc}

\usepackage{amssymb}
\usepackage{graphicx}
\usepackage{accents}
\usepackage[cmtip,all]{xy}
\usepackage{verbatim}
\usepackage{tikz-cd}
\usepackage{stmaryrd}
\usepackage{wasysym}
\usepackage{graphicx}
\usepackage{mathrsfs}
\usepackage{geometry}

\makeatletter
\DeclareRobustCommand{\cev}[1]{%
  \mathpalette\do@cev{#1}%
}
\newcommand{\do@cev}[2]{%
  \fix@cev{#1}{+}%
  \reflectbox{$\m@th#1\vec{\reflectbox{$\fix@cev{#1}{-}\m@th#1#2\fix@cev{#1}{+}$}}$}%
  \fix@cev{#1}{-}%
}
\newcommand{\fix@cev}[2]{%
  \ifx#1\displaystyle
    \mkern#23mu
  \else
    \ifx#1\textstyle
      \mkern#23mu
    \else
      \ifx#1\scriptstyle
        \mkern#22mu
      \else
        \mkern#22mu
      \fi
    \fi
  \fi
}

\makeatother

\DeclareMathAlphabet{\mathpzc}{OT1}{pzc}{m}{it}

\newtheorem{theorem}{Theorem}[section]

\theoremstyle{definition}
\newtheorem{definition}[theorem]{Definition}

\theoremstyle{remark}

\numberwithin{equation}{section}

 \newcommand{\virgolette}{``}

\newcommand{\slantone}[2]{{\raisebox{.1em}{$#1$}\left/\raisebox{-.1em}{$#2$}\right.}}
\newcommand*{\defeq}{\mathrel{\vcenter{\baselineskip0.5ex \lineskiplimit0pt
                     \hbox{\scriptsize.}\hbox{\scriptsize.}}}%
                     =}

\newcommand\asim{\mathrel{%
  \ooalign{\raise0.1ex\hbox{$\sim$}\cr\hidewidth\raise-0.8ex\hbox{\scalebox{0.9}{$\scriptstyle{x}$}}\hidewidth\cr}}}

\newcommand{\proj}[1]{\ensuremath{\mathbb{P}^{#1}}}

\newcommand{\mani}{\ensuremath{\mathpzc{M}}}
\newcommand{\manir}{\ensuremath{\mathpzc{M}_{red}}}
\newcommand{\stsheaf}{\ensuremath{\mathcal{O}_{\mathpzc{M}}}}
\newcommand{\stsheafred}{\ensuremath{\mathcal{O}_{\mathpzc{M}_{red}}}}

\newcommand{\beq}{\begin{equation}}
\newcommand{\eeq}{\end{equation}}
\newcommand{\bear}{\begin{eqnarray}}
\newcommand{\eear}{\end{eqnarray}}
\newcommand{\longhookrightarrow}{\ensuremath{\lhook\joinrel\relbar\joinrel\rightarrow}}

\begin{document}
\begin{flushright}
DISIT-2018\\
ARC-18-05
\par\end{flushright}

\title[]{\Large Superstring Field Theory,
\vspace{.2cm} \\
Superforms and Supergeometry}


\author{Roberto Catenacci}
\address{Dipartimento di Scienze e Innovazione Tecnologica - Università del Piemonte Orientale,
Via T. Michel 11, 15121, Alessandria, Italy} 
\address{Gruppo Nazionale di Fisica Matematica, InDAM, 
Piazzale Aldo Moro 5, 00185, Roma}
\address{Arnold Regge Center, Via P. Giuria, 10125, Torino, Italy}
\email{roberto.catenacci@uniupo.it}

\author{Pietro Antonio Grassi}
\address{Dipartimento di Scienze e Innovazione Tecnologica - Università del Piemonte Orientale, 
Via T. Michel 11, 15121, Alessandria, Italy} 
\address{INFN Sezione di Torino, Via P. Giuria 1, 10125 Torino}
\address{Arnold Regge Center, Via P. Giuria, 10125, Torino, Italy}
\email{pietro.grassi@uniupo.it}

\author{Simone Noja}
\address{Dipartimento di Scienze e Innovazione Tecnologica - Università del Piemonte Orientale, 
Via T. Michel 11, 15121, Alessandria, Italy} 
\address{INFN Sezione di Torino, Via P. Giuria 1, 10125 Torino}
\email{simone.noja@uniupo.it}







\begin{abstract}
Inspired by superstring field theory, we study differential, integral, and inverse forms and their mutual relations on a supermanifold from a sheaf-theoretical point of view. In particular, the formal distributional properties of integral forms are recovered in this scenario in a geometrical way. Further, we show how inverse forms \virgolette extend'' the ordinary de Rham complex on a supermanifold, thus providing a mathematical foundation of the Large Hilbert Space used in superstrings. Last, we briefly discuss how the Hodge diamond of a supermanifold looks like, and we explicitly compute it for super Riemann surfaces. 
\end{abstract}

\maketitle

\tableofcontents


\section{Introduction}

Supergeometry is a fascinating branch of mathematics that prompted from the physical motivation 
of describing fermionic degrees of freedom. As is well known since the first years of quantum mechanics, 
identical particles can appear in two types: bosons and fermions. They have different properties, but 
essentially their wave functions, describing the states of those particles, have to be either fully symmetrized 
under the exchange of two identical particles in the case of bosons, or fully anti-symmetrized in the case of fermions. Such a requirement is 
easily implemented by representing the fermions in terms of anticommuting variables, also said \emph{Grassmann 
variables} belonging to a superalgebra. This original motivation stemming from physics has given a strong impulse to study to the study of supergeometry, a context in which commuting and anticommuting variables can be treated on the same footing and described in a unified fashion. Nonetheless, further important developments were motivated by string theory and string field theory. 

In string theory, in order to include the fermionic physical degrees of freedom and also protecting the theory from unwanted tachyonic fields and stabilising the vacuum, one needs fermionic coordinates (either in the vector representation of Lorentz group, RNS formulation, or in the spinor representation, GS/pure spinor formulation). 
In this respect, the spacetime is enriched by these additional coordinates and the supergeometry starts playing a fundamental role. On one side, string theory needs the supergeometry formulation 
to define vertex operators, correlation functions and amplitudes, on the other side the geometry emerging from 
that embodies those anticommuting variables in the properties of supermanifolds. 

During the last years, several research articles \cite{Witten, WittenPerturbation} pointed out new important applications of supergeometry in the context of string theory. In particular, it has been observed that the correlation functions 
of vertex operators, after integrating over conformal fields, are special types of differential forms - known as 
\emph{integral forms} - on the supermoduli space of super Riemann surfaces. To complete the computation, one needs 
an integration on that supermoduli space, which proved to be a formidable hard problem as one has to confront with some typical supergeometric subtleties, as recently shown by Donagi and Witten in \cite{DonWit}. By the way, this kind of issues called for the definition of an integration theory on supermanifolds. This has been developed and it revealed new interesting features of differential forms: 
1) the differential forms on a supermanifold are characterized by two numbers: the \emph{form degree} and the 
\emph{picture number}, 2) the complex of superforms must be extended to integral forms. 
This is obtained by adding to the complex additional lines with fixed picture and variable form degree. 3) In general, picture-zero differential forms have no upper bound to their form degree, whilst integral forms, \emph{i.e.}\ those forms having maximal picture number, have no lower bound to their form degree (which can also be negative). Finally, differential forms with a generic picture number are unbounded from above and from below. In addition, at a given form number the forms with a non-maximal picture number span an infinite-dimensional space. 4) New differential operators can be defined in order to remove or to add picture to the differential forms. 

All these features are easily discussed in the context of conformal field theory where the calculations 
can be performed. Nevertheless, some of the computations have a geometrical origin and therefore 
these features can be translated in term of geometrical properties. For that purpose, we use a 
{\it sheaf-theoretical} approach to supermanifolds. Nonetheless, to keep our exposition as readable and concrete as possible, we will use as prototypical example for our considerations and constructions the \emph{projective superspaces} $\proj {n|m}$, whose supermanifold structure is non-trivial but easy-enough to allows us for explicit computations in order to identify the sheaves involved and make clear their sheaf-theoretical local-to-global nature. 
\noindent Also, some of the computational properties of integral forms are to be ascribed to their \emph{distributional} nature and therefore 
it is shown how analytical distributional properties and geometrical aspects fuse into a precise description. This also motivates the introduction of a new type of superforms, 
called negative-degree superforms or {\it inverse forms}, which have interesting properties. They play an essential role in the comparison between string theory and supergeometry. Indeed, in the string theory framework it is known how to enlarge the 
physical spectrum of states (called {\it Large Hilbert Space}) in order to gain a useful description of the BRST cohomology (vertex operator observables): in this paper we will show how this is achieved from a purely geometrical approach, shedding some light on the supergeometrical origin of concepts underlying string field theory. The 
Large Hilbert Space has new features that have never been considered in supergeometry revealing new interesting results. 
 
The main motivations of the present work is the translation into a rigorous mathematical framework of the
the properties of differential superforms, integral forms and inverse forms via sheaf theory. 
A future goal is to understand if the $A_\infty$-algebra appearing in super string field theory could show up also in the supergeometric context, possibly in a natural fashion.  

The plan of the paper is the following: in sec.~2, we revise some ingredients from physical perspective such 
as the beta-gamma ghost fields, their fermionization, their vertex operators and their OPE algebras. In addition, we recall some basics facts regarding distributions and how they have to be understood in the present context; finally, picture changing operator are preliminarily discussed here. In sec. 3 and sec. 4, we recall basic facts about supermanifolds and we introduce some of the natural sheaves (namely the tangent, the cotangent and the 
Berezinian sheaves) that can be defined over a supermanifold and that will enter our description. In sec. 5, we introduce a global definition of the sheaves of integral forms and related complex. In sec. 6, we introduce the new concept of negative-degree (\emph{a.k.a.} inverse forms) and their complex and we discuss the cohomology of Large Hilbert Space in two interesting instances. In sec. 7, some issues in higher odd dimensions are addressed and discussed. Finally, in sec. 8, using mostly \emph{Serre duality}, we briefly address the problem of attaching a \emph{Hodge diamond} to a complex supermanifold, by underlying the differences arising in comparison with the ordinary well-understood case: the relevant case of super Riemann surfaces is described in some details.

\section{The Large Hilbert Space, PCO's and New Superforms} \label{LHSPCO}

\noindent The ideas of the \textit{Large Hilbert Space} (LHS) and of the \textit{Picture
Changing Operators} (PCO) have been introduced in string theory \cite{FMS}, in order to
quantize the ghost fields associated to the superdiffeomorphisms on the
worldsheet. Nonetheless those ideas can be imported in the geometry of
supermanifolds and, as will be shown, lead to new interesting addition to the space of 
integral form. In particular, it will be shown that the space of
distributions such as the Dirac delta forms (local expressions for integral forms), used so far as
\textit{prototypes}  is not large enough and
it must be augmented to the full set of distributional forms with
compact support.

In the quantization of superstring theory (see \cite{POL} for a comprehensive
and complete review using the notation of the present section), one introduces
two sets of conformal fields with conformal weights $(2, -1)$ and $(3/2,
-1/2)$ needed to fix the local supersymmetry and worldsheet diffeomorphisms.
They are named \emph{ghost} and \emph{superghost fields} and denoted
by $(b(z),c(z))$ and $(\beta(z), \gamma(z))$, respectively. The first set is
made of anticommuting fields, while the second one by commuting
real fields. The quantization of the latter requires some additional care
since any function of the zero mode of $\gamma$ enters in the cohomology. Such a 
degree of freedom has the same properties of the differential $d\theta$ of the
worldsheet anticommuting local coordinate $\theta$ of the super Riemann surface in the local coordinate system $(z, \theta)$.

A powerful way to deal with the quantization of these fields is by performing a
\emph{fermionization} (see \cite{FMS}) by expressing the set
$(\beta(z),\gamma(z))$ in terms of two anticommuting fields $(\xi(z),\eta(z))$
(with conformal weight $(0,1)$) and one chiral boson $\phi(z)$ as follows
\begin{align}
&  \gamma(z)=\,:\eta(z)\,e^{\phi}(z):,\qquad\beta(z)=\,:\partial
\xi(z)\,e^{-\phi(z)}:,\nonumber\\
&  \delta(\gamma(z))=\,:e^{-\phi(z)}:,\qquad\;\delta(\beta(z))=\,:e^{\phi
(z)}:, \label{CIC}%
\end{align}
The colon notation, as usual, denotes the normal ordering in the products. 
In the second line, we have computed the Dirac delta functions of the fields
$\gamma(z)$ and $\beta(z)$ and it is not difficult to show that they indeed
satisfy the correct properties $\gamma\delta(\gamma)=0,$ $\gamma\delta
^{\prime}(\gamma)=-\gamma$ as for the usual Dirac distribution $\delta(x)$.
The tools needed are the elementary quantization techniques of conformal field
theory, reviewed in classical string theory manuals (\cite{GWS} and \cite{POL}).

The \textquotedblleft standard\textquotedblright\ Hilbert Space (or
\emph{Small Hilbert Space}, SHS henceforth) is identified with the Fock space
resulting from the quantization of the $(\eta,\xi)$ and $\phi$ conformal field
theories. In that space the zero mode of the field $\xi(z)$ is absent in the
expression (\ref{CIC}) and any operator built in terms of positive powers of
$\gamma,\beta$ and derivatives of $\delta(\gamma),\delta(\beta)$ can be easily
written without using the zero mode of $\xi$. For instance, we have
\begin{align}
&  \gamma^{p}=\frac{1}{(p-1)!}\,\eta\partial\eta\cdots\partial^{(p-1)}%
\eta\,e^{p\phi},\nonumber\\
&  \beta^{p}=\frac{1}{(p-1)!}\,\partial\xi\partial^{2}\xi\cdots\partial^{p}%
\xi\,e^{-p\phi},\nonumber \\
&  \delta^{(p)}(\gamma)=\partial\xi\cdots\partial^{p}\xi\,e^{-(p+1)\phi},\nonumber \\
&  \delta^{(p)}(\beta)=\eta\partial\eta\cdots\partial^{p-1}\eta\,e^{(p+1)\phi
}.
\end{align}
Switching to the usual language of supergeometry in a complex supermanifold of
dimension $1|1$, identifying $\gamma\sim d\theta$ and $\delta(\gamma
)\sim\delta(d\theta)$ and neglecting at the moment $dz,$ the expression
$\gamma^{p}$ belongs to $\Omega_{\mani}^{p;0}$ (the space of superforms of
zero picture), while $\delta^{(p)}(\gamma)$ belongs to $\Omega_{\mani}^{-p;1}$
(the space of integral forms see \cite{Belo1}). In the formulas (\ref{CIC}),
there are also the fields $\beta$ and $\delta(\beta)$: they are translated
into the geometric language as $\beta\sim\iota_{D}$, namely the interior
derivative, where $D=\partial_{\theta}$ and $\delta(\beta)\sim\delta(\iota
_{D})$ (notice that the interior derivative w.r.t.\ an odd vector field is an
even derivation, therefore the Dirac delta of $\iota_{D}$ is defined). To
invert the relation between the fields, we have
\begin{align}
\eta &  =\partial\gamma\,\delta(\gamma)=\partial\Theta(\gamma)\nonumber \\
\xi &  =\Theta(\beta),
\end{align}
where $\Theta$ is the Heaviside function, which can be given an integral
representation as
\[
\Theta({\mathcal{R}})=\lim_{\epsilon\rightarrow0^{+}}\Big(-i\int_{-\infty
}^{\infty}\frac{dt}{t-i\epsilon}\,\exp(-it{\mathcal{R}})\Big).
\]
for a given operator ${\mathcal{R}}$. Notice that, to represent completely the
field $\xi$ in terms of the original set of fields $\gamma,\beta$, one needs to
enlarge the space of distribution by considering also the Heaviside function.
Nevertherless, that distribution involves the field $\beta$, but apparently we
do not require the same enlargement also for $\gamma$. However, using
conformal field theory techniques\footnote{Namely, by using the OPE's
$\beta(z)\gamma(w)=\frac{1}{z-w}+(reg)$, we can indeed verify that $\xi
(z)\eta(w)=\frac{1}{z-w}+(reg)$.} one can show that, bringing the two
quantities $\Theta(\beta(z))$ and $\delta(\gamma(w))$ close to each other on
the worldsheet (represented here by the points $z$ and $w$ appearing in the
arguments of $\beta$ and of $\gamma$), we get the equation
\begin{equation}
\Theta(\beta(z))\,\delta(\gamma(w))=\frac{1}{\gamma(w)}+\ldots
\end{equation}
where the ellipsis stands for $O(z-w)$, namely those terms which are
polynomials in the difference of $z$ and $w$ and vanishing when $w\rightarrow
z$. This implies that the presence of the zero mode of $\xi$ allows us to
consider also the negative powers of $\gamma$. This fact has deep consequences
in string theory opening the possibility of constructing \emph{open superstring
field theory} \cite{Berko} and it has been used for proving \emph{Sen's conjecture}
\cite{Sen}.

We will show that also in the context of supermanifolds, we can consider a
\emph{Large Hilbert Space} (LHS), or better an enlarged space of forms
enriching the geometrical structures. For that purpose, by identifying
$\beta\sim\iota_{D}$ and $\gamma\sim d\theta$, we can compute the action of
the operator $\Theta(\iota_{D})$ on $\delta(d\theta)$ as follows
\begin{equation}
\Theta(\iota_{D})\delta(d\theta)=\Big(\lim_{\epsilon\rightarrow0^{+}}%
-i\int_{-\infty}^{\infty}\frac{dt}{t-i\epsilon}\,e^{-it{\mathcal{\iota}_{D}}%
}\Big)\delta(d\theta)=\lim_{\epsilon\rightarrow0^{+}}-i\int_{-\infty}^{\infty
}\frac{dt}{t-i\epsilon}\delta(d\theta-it)=\frac{1}{d\theta}%
\end{equation}
This new relation shows that, by allowing for the differential operator
$\Theta(\iota_{D})$, we are forced to consider also negative powers of
$d\theta$ along the same ideas pursued in string theory. Using the operator
$\Theta(\iota_{D})$, we are able to map the integral forms complex
$\Omega_{\mani}^{p;1}$ 
into the new complex of
superforms with negative degree.

The generalisation to derivatives of delta functions is
\begin{equation}
\Theta(\iota_{D})\delta^{(n)}(d\theta)=(-1)^{n}n!\frac{1}{(d\theta)^{n}}\,.
\end{equation}
Therefore, we have a map
\begin{equation}
\Theta(\iota_{D}):\Omega_{\mani}^{p;1}\longrightarrow\Omega_{\mani}^{p-1;0}%
\end{equation}
for $p\in\mathbb{Z}$ and $p\leq1$, and where $\Omega_{\mani}^{p-1;0}$ for $p-1 \leq 0$ denotes
the space of superforms with inverse powers of $d\theta$. Note that we have to
take into account that the derivatives $\delta^{(p)}(d\theta)$ are required to
be anticommuting quantities in order to be able to build full-fledged complex
of integral forms and the corresponding top forms. In the same way, the
distribution $\Theta(\iota_{D})$ is also an anticommuting differential
operator acting on the space of integral forms. Therefore, the action of
$\Theta$ on $\delta(d\theta)$ yields a commuting quantity, namely
$(d\theta)^{-1}$, which is consistent with the algebraic properties.

However from the analytic point of view, we have to clarify an important issue. As
is well known, the distributions also emerge by introducing the famous
$i\epsilon$-prescription and using the formula (Sokhotski-Plemelj theorem
\cite{schwartz})
\begin{equation} \label{kpt}
\lim_{\epsilon\rightarrow0^{+}}\frac{1}{x-x_{0}\pm i\epsilon}=\mathrm{p.v.}%
\Big(\frac{1}{x-x_{0}}\Big)\mp i\pi\delta(x-x_{0})
\end{equation}
where $x,x_{0}$ are defined on $\mathbb{R}$, { }$\mathrm{p.v.}$ stands for
\emph{principal value} and it is defined as usual as
\begin{equation}
\Big\langle\mathrm{p.v.}\Big(\frac{1}{x}\Big),f(x)\Big\rangle=\lim
_{\epsilon\rightarrow0}\Big(\int_{\epsilon}^{\infty}\frac{f(x)dx}{x}%
+\int_{-\infty}^{-\epsilon}\frac{f(x)dx}{x}\Big)\,.
\end{equation}
for any test function $f(x)$ with compact support. The integral representation
of the Dirac delta function $\delta(x)$ used in the literature contains a
$\frac{1}{2\pi}$ factor bringing the factor $\pi$ in the above expression. If
we would like to use the same expression for $x\leftrightarrow d\theta$ and
$x_{0}\leftrightarrow0$, taking into account that $\delta(d\theta)$ is an
anticommuting operator, we have
\begin{equation}
\lim_{\epsilon\rightarrow0^{+}}\frac{1}{d\theta\pm i\epsilon}=\mathrm{p.v.}%
\Big(\frac{1}{d\theta}\Big)\pm\frac{i}{2}\Pi\delta(d\theta)\label{LHSD}%
\end{equation}
where $\Pi$ is the parity changing functor described in the following section
and the inverse power of $d\theta$ is considered as the distribution
$\mathrm{p.v.}(1/d\theta)$, which is a compact support distribution. Using the
$\Pi$ funtor, we correctly take into account the algebraic properties (the
number $\pi$ disappeared because our integral representation for Dirac delta
function is
\begin{equation}
\delta(d\theta)=\int_{-\infty}^{\infty}\exp\Big(itd\theta\Big)dt
\end{equation}
and it does not have the $1/2\pi$ factor in it. The two terms in the
r.h.s.\ of (\ref{LHSD}) are two distributions with different characteristics
and different degrees. In particular they belong to $\Omega_{\mani}^{-1;0}$
and $\Omega_{\mani}^{0;1}$. It is worthwhile noting that the transformation
properties of both expressions in the r.h.s. of the equations, under change of
patches are exactly the same. This point will be completely elucidated in the
forthcoming sections where a coordinated-free definition of the objects
considered in this section will be provided.

Still working on a local set of coordinates, we can multiply both sides of
(\ref{LHSD}) by $\theta$ to get
\begin{equation}
\lim_{\epsilon\rightarrow0^{+}}\frac{\theta}{d\theta\pm i\epsilon
}=\mathrm{p.v.}\Big(\frac{\theta}{d\theta}\Big)\mp\frac{i}{2}\Pi\theta
\delta(d\theta)=\alpha^{(-1|0)}\mp\frac{i}{2}\Pi\mathbb{Y}^{(0|1)}%
\end{equation}
where we have defined the two quantities
\begin{equation}
\alpha^{(-1|0)}\defeq {\rm p.v.}\Big(\frac{\theta}{d\theta}\Big)\,,\hspace
{2cm}\mathbb{Y}^{(0|1)}\defeq\theta\delta(d\theta)\,.
\end{equation}
called \textit{trivializer} and \textit{Picture Changing Operator} (PCO),
respectively. The name of the first one is due to its property
\begin{equation}
d\left(  \lim_{\epsilon\rightarrow0^{+}}\frac{\theta}{d\theta\pm i\epsilon
}\right)  =d\left[  \mathrm{p.v.}\Big(\frac{\theta}{d\theta}\Big)\right]  \pm
i\Pi d\Big(\theta\delta(d\theta)\Big)=1
\end{equation}
which implies that $d\alpha^{(-1|0)}=1$ and $d\mathbb{Y}^{(0|1)}=0$. Indeed, 
$\alpha^{(-1|0)}$ is the trivializer of the usual odd differential operator
$d$. The second operator is imported from string theory where it plays a
fundamental role for constructing the amplitudes. It is $d$-closed but it is
not exact and, as such, it belongs to the de Rham cohomology group
$H_{dR}^{0;1}(\mathpzc{M})$ of the supermanifold.

The generalisation to higher powers is:
\begin{equation}
\lim_{\epsilon\rightarrow0^{+}}\frac{\theta}{d\theta^{2}\pm i\epsilon
}=\mathrm{f.p.}\Big(\frac{\theta}{(d\theta)^{2}}\Big)\mp\frac{i}{2}\Pi
\theta\delta^{\prime}(d\theta)
\end{equation}
where $\mathrm{f.p.}$ is the Hadamard finite part of the integral defined as
\begin{equation}
\Big\langle\mathrm{f.p.}\left(  \frac{1}{x^{2}}\right)  ,f(x)\Big\rangle=\lim
_{\epsilon\rightarrow0}\left[  \int_{-\infty}^{-\epsilon}\frac{f(x)dx}{x^{2}%
}+\int_{\epsilon}^{\infty}\frac{f(x)dx}{x^{2}}-\frac{2f(0)}{\epsilon}\right]
\end{equation}
for any test function $f(x)$ with compact support.

Let us consider the two elements of $\Omega_{\mani}^{0;1}$ given by
\begin{equation}
\mathbb{Y}^{(0|1)}=\theta\delta(d\theta)\,,\quad\quad\widetilde{\mathbb{Y}%
}^{(0|1)}=(dz-\theta d\theta)\delta^{\prime}(d\theta).\label{PPA}%
\end{equation}
They differ by an exact term, as can be readily checked. We can in the same
way define the operators
\begin{equation}
\alpha_{\epsilon}^{(-1|0)}=\frac{\theta}{d\theta\pm i\epsilon}\,,\qquad
\widetilde{\alpha}_{\epsilon}^{(-1|0)}=\frac{(dz-\theta d\theta)}{d\theta
^{2}\pm i\epsilon}\,,\label{PPE}%
\end{equation}
satisfying the equations
\begin{equation}
d\alpha_{\epsilon}^{(-1|0)}=1\,,\quad d\widetilde{\alpha}_{\epsilon}%
^{(-1|0)}=1\,,\quad\widetilde{\alpha}_{\epsilon}^{(-1|0)}-{\alpha}_{\epsilon
}^{(-1|0)}=d\Omega^{(-2|0)}.\label{PPF}%
\end{equation}
The Large Hibert Space is spanned by the superforms
\begin{equation}
(d\theta)^{p},(d\theta\pm i\epsilon)^{-p-1}, \qquad p\geq0\,.\label{PPG}%
\end{equation}
Since there are two regularizations for the inverses of $d\theta$ associated
to the two signs $\pm i\epsilon$, the Hilbert Space is not isomorphic to the
original one. \\
Equivalently, the Large Hilbert Space is spanned by the superforms 
\bear
(d\theta)^{p}, \, \delta^{(p)} (d\theta) , \,\mbox{f.p.}\left ( \frac{1}{d\theta^{p}} \right ), \qquad  p \geq 0.
\eear

\noindent The Heaviside step operator $\Theta(\iota_{D}) $ enters the
definition of another type of PCO that is given by (see \cite{Belo1}):%
\begin{equation}
\mathbb{Z}_{D}=\left[  d,\Theta\left(  \iota_{D}\right)  \right]  .
\label{Zeta}%
\end{equation}
which depends on the choice of the vector field $D$. Note that, being $d$ an
odd differential and $\Theta(\iota_{D})$ an odd operator,
the PCO $\mathbb{Z}_{D}$ is a an even operator.

Acting on $\mathbb{Y}^{(0|1)}$ we get:
\begin{equation}
\Theta(\iota_{D})\mathbb{Y}^{(0|1)}=\frac{\theta}{d\theta}\,,\hspace
{1cm}\mathbb{Z}^{(0|-1)}\mathbb{Y}^{(0|1)}=d\left[  \Theta(\iota
_{D})\mathbb{Y}^{(0|1)}\right]  =d\left(  \frac{\theta}{d\theta}\right)
=1\,.\label{PPB}%
\end{equation}


The $\mathbb{Z}_{D}$ operator is in general not invertible but it is possible
to find a \textit{non unique} operator $\mathbb{Y}$ such that $\mathbb{Z}%
\circ\mathbb{Y}$ is an isomorphism in de Rham cohomology \textit{i.e.}\ the
cohomology of the $d$ operator described above. These operators are the called
\textit{Picture Raising Operators. }The operators of type $\mathbb{Y}$ are non
trivial elements of the de Rham cohomology.

We apply a PCO of type $\mathbb{Y}$ on a given form by taking the graded wedge
product: given $\omega$ in $\Omega_{\mani}^{p;q}$, we have:
\begin{equation}
\omega\overset{\mathbb{Y}}{\longrightarrow}\omega\wedge{\mathbb{Y}}\in
\Omega_{\mani}^{{p;q+1}}\label{PCOc}%
\end{equation}
If $q=m$, then $\omega\wedge{\mathbb{Y}}=0$. In addition, if $d\omega=0$ then
$d(\omega\wedge{\mathbb{Y}})=0$, and if $\omega\neq dK$ then it follows that
also $\omega\wedge Y\neq dU$ where $U$ is a form in $\Omega_{\mani}^{p-1;q+1}%
$. So, given an element of the cohomogy $H_{DR}^{p;q}(\mathpzc{M})$, the new
form $\omega\wedge Y$ is an element of $H_{dR}^{p;q+1}(\mathpzc{M})$. The
${\mathbb{Y}}$ and ${\mathbb{Z}}$ operators give an isomorphism in de Rham
cohomology:%
\begin{equation}
H_{dR}^{p;0}(\mathpzc{M})\cong H_{dR}^{p;m}(\mathpzc{M})\label{deli}%
\end{equation}
Incidentally, this imply that $H_{dR}^{p;0}(\mathpzc{M})=\left\{  0\right\}  $
for $p>n\ $and this means that the de Rham cohomology of superforms cannot
capture informations on the odd dimensions \cite{Deligne}, \cite{VorGeom}.


We can build explicitely a \textit{left} inverse for the $\Theta$ operator
that it is called $\eta_{0}$. From \eqref{kpt}
\begin{equation}
\lim_{\epsilon\rightarrow0}\frac{1}{2i}\Big(\frac{1}{x-i\epsilon}-\frac
{1}{x+i\epsilon}\Big)=\lim_{\epsilon\rightarrow0}\frac{\epsilon}%
{x^{2}+\epsilon^{2}}=\delta(x)\label{newA}%
\end{equation}
The expression on the left hand side can be rewritten as follows (using the
translation operator $e^{i\epsilon\partial_{x}}$)
\begin{equation}
\lim_{\epsilon\rightarrow0}\frac{1}{2i}\Big(e^{i\epsilon\partial_{x}%
}-e^{-i\epsilon\partial_{x}}\Big)\frac{1}{x}=\lim_{\epsilon\rightarrow0}%
\sin(\epsilon\partial_{x})\frac{1}{x}=\delta(x)\label{newB}%
\end{equation}
Let us consider now the formal series $f(x)=\sum_{n=-\infty}^{\infty}%
c_{n}x^{n}$; for each single term with $n>0$ we have
\begin{equation}
\lim_{\epsilon\rightarrow0}\sin(\epsilon\partial_{x})x^{n}=\lim_{\epsilon
\rightarrow0}\frac{1}{2i}\Big((x+i\epsilon)^{n}-(x-i\epsilon)^{n}%
\Big)=0\,,\label{newC}%
\end{equation}
As for the negative powers $n<0$, we can set $x^{-n}=(-)^{n}/(n-1)!\left.
\partial_{\alpha}^{\left(  n\right)  }(x+\alpha)^{-1}\right\lfloor _{\alpha
=0}$, then we have
\begin{equation}
\lim_{\epsilon\rightarrow0}\sin(\epsilon\partial_{x})x^{-n}=\frac{(-)^{n}%
}{(n-1)!}\left.  \partial_{\alpha}^{\left(  n\right)  }\left(  \lim
_{\epsilon\rightarrow0}\sin(\epsilon\partial_{x})\frac{1}{(x+\alpha)}\right)
\right\lfloor _{\alpha=0}=\frac{(-)^{n}}{(n-1)!}\delta^{(n)}(x)\label{newD}%
\end{equation}
Translating for the even superform $d\theta$ we define,
\begin{equation}
\eta_{0}\defeq\Pi\lim_{\epsilon\rightarrow0}\sin(\epsilon\iota_{D}%
)\label{etazero}%
\end{equation}
where again we have used the parity-changing functor $\Pi$ in order to assign
the correct parity to the operator, consistently with the properties of
$d\theta$ and of $\delta(d\theta)$. We can check that:
\begin{equation}
\eta_{0}\delta(d\theta)=0\,,
\end{equation}
and
\begin{equation}
\eta_{0}\Theta(\iota_{D})\delta(d\theta)=\eta_{0}\left(  \frac{1}{d\theta
}\right)  =\delta(\,d\theta)\,,~~~~~~\Theta(\iota_{D})\eta_{0}\delta\,\left(
d\theta\right)  =0\,.\label{etazeroC}%
\end{equation}
Then, $[\eta_{0},\Theta(\iota_{D})]=1$ which is what we want.

\section{Elements of Supermanifolds}

\noindent In this section we shall give the most important definitions in the theory of supermanifolds, in order to set some notation and terminology. For a more complete introduction to the theory of supermanifolds via algebraic geometry we suggest the reader to refer to the deep treatment given by Manin in \cite{Manin}, some details of which have been recently spelled out in \cite{NojaPhD}.

As a general setting, we work in the (super) analytic category and we take our ground field to be the complex numbers $\mathbb{C}$.

Our main characters will be \emph{complex supermanifolds}. In general, a complex supermanifold of dimension $n|m$ is a locally ringed space $(\mani, \stsheaf)$, where $\mani$ is a topological space and $\stsheaf = \mathcal{O}_{\mani,0}\oplus \mathcal{O}_{\mani,1}$ is a sheaf of supercommutative algebras over $\mathbb{C}$ on $\mani$, called the \emph{structure sheaf} of the supermanifold, such that the following conditions are satisfied:
\begin{enumerate}
\item the pair $(\mani, \stsheafred)$, where $\stsheafred \defeq \stsheaf / \mathcal{J}_\mani$, for $\mathcal{J}_\mani \defeq \mathcal{O}_{\mani,1} \oplus \mathcal{O}_{\mani, 1}^{\otimes2}$, is a \emph{complex manifold} of dimension $n$. The pair $(\mani, \stsheafred)$ is called the \emph{reduced space} of the supermanifold $(\mani, \stsheaf)$;
\item the quotient $\mathcal{J}_\mani / \mathcal{J}_\mani^2$ is a locally-free sheaf of $\stsheafred$-modules of rank $0|m$, and it is called the \emph{fermionic sheaf} and denoted by $\mathcal{F}_\mani$; 
\item the structure sheaf $\stsheaf$ is \emph{locally} isomorphic to the \emph{exterior algebra} $\bigwedge^\bullet \mathcal{F}_\mani$ over $\stsheafred$, seen as a superalgebra. 
\end{enumerate}
For the sake of brevity, we will denote a supermanifold by $\mani$ and its reduced space by $\manir$. Also, following \cite{Manin}, it is worth noticing that since $\mathcal{F}_\mani$ is a purely odd sheaf, it would be more appropriate to write $Sym^\bullet \mathcal{F}_\mani$, instead of $\bigwedge^\bullet \mathcal{F}_\mani$, as we will do later on in this paper. 

Before we go on we make some comments on this definition. First of all we remark that the first condition mentioned above corresponds, for any supermanifold $\mani$, to the existence of a morphism of supermanifolds $\iota : \manir \rightarrow \mani$ such that $\iota$ is a pair $\iota \defeq (\iota, \iota^\sharp)$, with $\iota: \mani \rightarrow \mani$ the identity on the underlying topological space and $\iota^\sharp : \stsheaf \rightarrow \stsheafred$ is the quotient map by $\mathcal{J}_\mani$, the \emph{sheaf of ideals} formed by all of the nilpotents. Loosely speaking, this tells that the reduced manifold arises by {setting all of the nilpotents in $\stsheaf$ to zero}. More precisely, a more invariant formulation is that to any supermanifold is attached a short exact sequence as follows
\bear \label{defses}
\xymatrix{
0 \ar[r] & \mathcal{J}_\mani \ar[r] & \mathcal{O}_{\mani} \ar[r]^{\iota^\sharp\; } & \mathcal{O}_{\manir} \ar[r] & 0, 
}
\eear
that tells that the structure sheaf $\mathcal{O}_\mani$ of a supermanifold is an \emph{extension} of $\mathcal{O}_{\manir}$ by $\mathcal{J}_\mani$. In view of this, a very important question in the theory of supermanifolds is whether the defining short exact sequence \eqref{defses} is \emph{split} or not, \emph{i.e.}\ whether there exists a morphism of supermanifolds $\pi : \mani \rightarrow \manir$ given by a pair $(\pi , \pi^\sharp )$ with $\pi : \mani \rightarrow \mani$ being again the identity on the underlying topological space and $\pi^\sharp : \mathcal{O}_{\manir} \rightarrow \mathcal{O}_\mani$ a splitting morphism - called a \emph{projection} - such that $ \pi^\sharp \circ \iota^\sharp = id_{\stsheaf}.$ In case the splitting morphism $\pi : \mani \rightarrow \manir$ exists, then $\mani$ is said a \emph{projected} supermanifold, otherwise is said a \emph{non-projected}.
In this paper we will not be concerned with the subtleties related to non-projected supermanifolds - which yield complicated problems in the theory of complex supermanifolds that deeply affects the computation of amplitudes in superstring theory \cite{DonWit} -, by the way the interested reader is advised to refer to the recent \cite{CNR} and \cite{PiGeo} to get an idea about the phenomenology related to these kind of supermanifolds.\\
The third condition in the definition of a supermanifold is often briefly referred in short by saying that a complex supermanifold of dimension ${n|m}$ is locally isomorphic to the superspace $\mathbb{C}^{n|m}$. Actually, more precisely, this third condition is the fundamental request that the structure sheaf $\stsheaf$ is \emph{locally} \emph{freely-generated} by linear independent sections, we will denote them by $(x_1, \ldots, x_n, \theta_1, \ldots, \theta_m)$. These are subjected to supercommutativity only: this implies that, locally, every section $s$ in $\stsheaf$ can be uniquely represented by a power expansion in the theta's, that is 
\bear
s (x, \theta) = s_0 (x) + s_i (x) \theta^i + s_{ij} (x) \theta^i \theta^j \ldots, 
\eear
in an open set ${U}\subseteq \mani$ and where $s_0, s_i, s_{ij}, \ldots $ are sections in $\stsheafred$ over ${U}$, \emph{i.e.}\ holomorphic functions over $\mathcal{U}$. It is crucial to note that since the theta's are nilpotent, this power expansion has a finite number of terms, actually $2^m$ at most. \\ 
A projected supermanifold whose structure sheaf is given itself by an exterior algebra is said to be \emph{split}. \vspace{.3cm}

\noindent The most important class of split complex supermanifolds is given by (complex) \emph{projective superspaces} $\proj {n|m} \defeq (\proj n, \mathcal{O}_{\proj {n|m}} )$, where $\mathcal{O}_{\proj {n|m}} \defeq Sym^\bullet \left (\mathbb{C}^m \otimes_{\mathbb{C}} \mathcal{O}_{\proj n}(-1) \right ))$, that is, more extensively, 
\bear
\mathcal{O}_{\proj {n|m}} \defeq \bigoplus_{k \, even} \bigwedge^k \mathcal{O}_{\proj {n}} (-1)^{\oplus m} \oplus \bigoplus_{k \, odd} \bigwedge^k \mathcal{O}_{\proj {n}} (-1)^{\oplus m}.
\eear
\noindent 
When it comes to sheaf-theoretic constructions, projective superspaces $\proj {n|m}$ represent a particularly suitable class of examples as they allow for an immediate and easy \emph{local-to-global} and \emph{global-to-local} correspondence, and as such they will be extensively used throughout the paper.\\
A projective superspace $\proj {n|m}$ has a straightforward local description by patching affine charts. Since the underlying reduced manifold of $\proj {n|m}$ is just $\proj n$, it has a covering $ \{U_i\}_{i=0,\ldots, n}$ made by $n+1$ open sets $U_i$, each characterized by the usual condition on the homogeneous coordinates. These open sets, in turn, make up $n+1$ affine supermanifolds $\tilde U_i \cong \mathbb{C}^{n|m}$ with $\tilde U_i \defeq (U_i , \mathbb{C}[z_{\ell i} , \theta_{\kappa i}] )$, for $i= 0, \ldots, n$ and $\ell \neq i $, $\kappa = 1, \ldots, m$, which cover $\proj {n|m}$. \\
In the intersections $U_i \cap U_j $ for $0 \leq i < j \leq n+1$ transition functions reads
\begin{align} \label{transi}
& \ell \neq i : \qquad  \qquad \qquad \; z_{\ell j} = \frac{z_{\ell i}}{z_{j i}},   \nonumber \\
& \ell = i : \qquad \qquad \qquad \; z_{i j} = \frac{1}{z_{ji}} , \nonumber \\
& \kappa =  1, \ldots, m: \qquad \quad {\theta_{\kappa j}} = \frac{\theta_{\kappa i}}{z_{ji}},
\end{align}
and this gives an atlas for $\proj {n|m}$. \\
The interested reader can find a detailed treatment of the supergeometry of projective superspaces in the recent \cite{CN}.

\section{Locally-Free Sheaves on Supermanifolds: Tangent, Cotangent and Berezinian Sheaves}

\noindent Now that we have introduced what a supermanifold is, let us see what can be defined on it. For our purposes, one of the most important and useful concept is the one of {locally-free sheaf}, that will completely replace the cumbersome notion of super vector bundle \cite{Manin}, \cite{NojaPhD}. \\
Given a supermanifold $\mani$, a \emph{locally-free sheaf} $\mathcal{G}$ of rank $p|q$ on $\mani$ is simply a sheaf of $\stsheaf$-modules which is locally-isomorphic to $\stsheaf^{\oplus p} \oplus \left (\Pi \stsheaf \right )^{\oplus q}$, where $\Pi \mathcal{O}_{\mani}$ is structure sheaf of the supermanifold having reversed parity.\\
In particular, an \emph{even invertible sheaf} $\mathcal{L}_{ev}$ on $\mani$ is a rank $1|0$ locally-free sheaf of $\stsheaf$-modules, and, likewise, an \emph{odd invertible sheaf} $\mathcal{L}_{odd}$ on $\mani$ is a rank $0|1$ locally-free sheaf of $\stsheaf$-modules. This means that, locally, one has $\mathcal{L}_{ev} \lfloor_{U} \cong \stsheaf \lfloor_{U}$ and $\mathcal{L}_{odd} \lfloor_{U} \cong \Pi \stsheaf \lfloor_{U}$ for $U$ an open set of $\mani$. \\
Note that in general, exactly as in the ordinary context, defining a locally-free sheaf $\mathcal{G}$ of a certain rank on a supermanifold $\mani$, amounts to give an open covering of $\mani$, call it $\{{U}_i \}_{i \in I}$, and the transition functions $\{ g_{ij} \}_{i,j \in I}$ between two local frames - call them $e_{{U}_i} $ and $e_{{U}_j}$ - in the intersections ${U}_{i} \cap {U}_j$ for $i, j \in I$, so that $e_{{U}_i } = g_{ij} e_{{U}_j}$. In this fashion, one finds the usual correspondence $\mathcal{G} \leftrightarrow \left ( \{ {U}_i \}_{i \in I}, \{ g_{ij}\}_{i,j \in I} \right ),$ where if $\mathcal{G}$ has rank $p|q$ then $g_{ij}$ is a $GL_{p|q}$ transformation taking values in $\mathcal{O}_\mani ({U}_i \cap {U}_j))$. \\
In the case we are considering, say, an even invertible sheaf, this corresponds to transition functions $g_{ij}$ taking values into $\left (\mathcal{O}_{\mani}^\ast \right )_0 \cong \mathcal{O}^\ast_{\mani,0}$, the sheaf of \emph{non-vanishing} sections of the structure sheaf. This is so as the transformation $g_{ij}$ needs to be \emph{invertible} and a \emph{parity-preserving one}: indeed the frames have well-defined parity that get preserved under a change of coordinates. This bears an important consequence: $\mathcal{O}^\ast_{\mani,0}$ is a sheaf of \emph{abelian groups}, so that we are allowed to consider its cohomology groups, without confronting the issues related to the definition of non-abelian cohomology (the full sheaf $\mathcal{O}^\ast_\mani$ is indeed \emph{not} a sheaf of abelian groups). Notice that in order to define an even invertible sheaf, the transition functions have to be $1$-cocycles valued in the sheaf $\mathcal{O}^\ast_{\mani,0}$, so that one has the super-analog of the usual correspondence between the \emph{even Picard group} $\mbox{Pic}_0(\mani) $ - which is the group of the isomorphism classes of even invertible sheaves on $\mani$ - and the cohomology group $H^1 (\mani, \mathcal{O}_{\mani,0}^\ast)$, see \cite{CN} or \cite{ManinNC} for details.\\
Clearly, the classification and the related moduli problem for higher rank sheaves is much more difficult, and, just like in the ordinary theory, being $GL_{p|q} (\stsheaf)$ non-commutative, the set $H^1 (\mani, GL_{p|q} ( \mathcal{O}_\mani))$ is not endowed with a group structure, but it is just a pointed-set instead, whose identity is usually taken to be the trivial rank $p|q$ sheaf. By the way we shall not worry about these subtleties in what follows, as we will not be interested into a classification but just into identifying certain sheaves instead, so it will be enough to look at the specific form of the transition functions.\vspace{.5cm}

In comparison with the usual commutative geometric context, there is at least one more important operation one can do on a locally-free sheaf $\mathcal{E}$ on a supermanifold $\mani$, that is to \emph{reverse its parity}. Indeed, let $\mathcal{E}$ be a rank $p|q$ sheaf, which is freely-generated in an open set $U$ as follows 
\bear
\mathcal{E}\lfloor_{U} \cong \mathcal{O}_{\mani} \lfloor_{U} \cdot \{ e^{(0)}_1, \ldots, e^{(0)}_p | e^{(1)}_1, \ldots, e^{(1)}_q  \},
\eear 
where $\{ e^{(0)}_1, \ldots, e^{(0)}_p | e^{(1)}_1, \ldots, e^{(1)}_q  \}$ is a local frame of generators over $U$, the upper indices refer to the parity of the generators and, as a general convention in this paper, the even generator will be written in the first place. Then, acting with the parity-changing functor yields a rank $q|p$ locally-free sheaf, we call it $\Pi \mathcal{E}$. This is freely-generated over $U$ by 
\bear
\Pi \mathcal{E}\lfloor_{U} \cong \mathcal{O}_{\mani} \lfloor_{U} \cdot \{ \pi e^{(1)}_1, \ldots, \pi e^{(1)}_q | \pi e^{(0)}_1, \ldots, \pi e^{(0)}_p  \},
\eear
where we have denoted with $\pi e^{(0)}_i$ and $\pi e^{(1)}_j$ the images of the generators $e^{(0)}_i$ and $e^{(1)}_j$ for $i=1, \ldots, p$ and $j = 1, \ldots, q$ under action of the parity-changing functor $\Pi$. That is, in other words, 
\bear
\left (
 \begin{array}{l}
e^{(0)}_1  \\
\vdots \\
e^{(0)}_p \\
\hline
e^{(1)}_1\\
\vdots \\
e^{(1)}_q
\end{array}
\right ) \xymatrix{ \ar@{|->}[rr]^{\Pi} &&
}
\left (
 \begin{array}{l}
\pi e^{(1)}_1  \\
\vdots \\
\pi e^{(1)}_q \\
\hline
\pi e^{(0)}_1\\
\vdots \\
\pi e^{(0)}_p
\end{array}
\right )\eear
where as for the parity one has $| e^{(p)}_\ell |=  p $ and $|\pi e^{(p)}_\ell |= ( p+1 )\mbox{mod} 2$, for $p \in \mathbb{Z}_2$ and any $\ell.$ \\
Notice that, as observed above, given a covering $\{U_i \}_{i \in I}$ of a complex manifold $\mani$, one can present a locally-free sheaf $\mathcal{E}$ by giving its transition functions $g_{ij}(\mathcal{E}) : \mathcal{E}(U_i) \lfloor_{U_i \cap U_j} \rightarrow \mathcal{E} (U_j) \lfloor_{U_i \cap U_j} $ in the intersections ${U}_i \cap {U}_j$, so that the sheaf $\mathcal{E} $ is identified by the pair $(\{U_i\}_{i \in I}, \{ g_{ij}(\mathcal{E}) \}_{i, j \in I} )$. From this point of view it is easy to identify the sheaf $\Pi \mathcal{E}$. We denote by $M (g_{ij}(\mathcal{E}))$ the transition matrix related to $g_{ij} \mathcal (\mathcal{E})$, so that one can write in general  
\bear
M (g_{ij}(\mathcal{E}))= \left ( 
\begin{array}{c|c}
A_{p\times p} & B_{p\times q} \\
\hline 
C_{q\times p} & D_{q \times q}
\end{array}
\right ) \in GL_{n|q} (\mathcal{O}_\mani (U_i \cap U_j))
\eear 
for some invertible matrices $A_{p\times p } \in GL_p (\mathcal{O}_{\mani,0} (U_i \cap U_j)), D_{q \times q } \in GL_{q} (\mathcal{O}_{\mani, 0} (U_i \cap U_j))$ and some matrices $B_{p \times q} \in Mat_{p\times q} ( \mathcal{O}_{\mani,1} (U_i \cap U_j))$, $C_{q \times p} \in Mat_{q\times p} (\mathcal{O}_{\mani, 1} (U_i \cap U_j))$. The transition functions $\{ g_{ij} (\Pi \mathcal{E}) \}_{i,j \in I}$ of the sheaf $\Pi \mathcal{E}$ are then immediately recovered from the $M (g_{ij}(\mathcal{E}))$'s via the \emph{parity-transpose} operation:
\bear
M (g_{ij}(\Pi \mathcal{E})) = M (g_{ij}(\mathcal{E}))^{\Pi } \defeq \left ( 
\begin{array}{c|c} D_{q \times q} & C_{q\times p} \\
\hline 
B_{p\times q} & A_{p\times p}
\end{array}
\right ) \in GL_{q|p} (\mathcal{O}_\mani (U_i \cap U_j)),
\eear
in other words one finds that given $\mathcal{E} \leftrightarrow ( \{U_i\}_{i \in I}, \{ g_{ij}(\mathcal{E}) \}_{i, j \in I} )$, then the sheaf $\Pi \mathcal{E}$ is simply given by $\Pi \mathcal{E} \leftrightarrow (\{U_i\}_{i \in I}, \{ g^\Pi_{ij}(\mathcal{E}) \}_{i, j \in I})$, where we have indicated with $g_{ij}^\Pi (\mathcal{E})$ the parity-transpose operation on the transition functions, as explained above.\vspace{.5cm}

Let us see some instances of what explained above via some concrete examples. In particular, let us consider two locally-free sheaves that can be naturally defined on any supermanifold, the \emph{tangent sheaf} $\mathcal{T}_\mani$ and the \emph{cotangent sheaf} $\mathcal{T}^\ast_\mani$.\\
The tangent sheaf of $\mani$ is defined, as usual, as the sheaf of {superderivation} on $\mani$, where for a superalgebra $A$ a \emph{superderivation} is a homogeneous $k$-linear maps $D : A \rightarrow A$ of parity $|D| \in \mathbb{Z}_2$ that satisfies the $\mathbb{Z}_2$-graded Leibniz rule: 
\bear
D(a \cdot b) = D(a) \cdot b + (-1)^{|D| |a|} a \cdot D(b),
\eear 
for any $a \in A$ homogeneous of parity $|a|$ and any $b \in A.$ In particular, on the complex superspace $\mathbb{C}^{n|m}$ having coordinates $( x_1, \ldots, x_n | \theta_1, \ldots, \theta_m )$, the superderivations of the structure sheaf $\mathcal{O}_{\mathbb{C}^{n|m}}$ are written as $( \partial_{x_1}, \ldots, \partial_{x_n} | \partial_{\theta_1}, \ldots, \partial_{\theta_n}),$ where the $\{ \partial_{x_i}\}_{i=1, \ldots, n}$ are the \emph{even} superderivations and the $\{\partial_{\theta_j} \}_{j=1, \ldots m}$ are the \emph{odd} superderivations and it is an early result in the theory of supermanifolds due to Leites (see \cite{Deligne}) that the $\mathcal{O}_{\mathbb{C}^{p|q}}$-module of the $\mathbb{C}$-linear superderivations is \emph{free} and has dimension $n|m$ with basis given indeed by $\{ \partial_{x_1}, \ldots, \partial_{x_n} | \partial_{\theta_1}, \ldots, \partial_{\theta_m} \}$. It follows that, since a complex supermanifold $\mani$ of dimension $n|m$ is by definition locally isomorphic to $\mathbb{C}^{n|m}$, the $\mathcal{O}_{\mathbb{C}^{n|m}}$-module of superderivations of the structure sheaf $\stsheaf$ is actually a locally-free sheaf of $\stsheaf$-modules of rank $n|m$ and we denote if by $\mathcal{T}_\mani$ and refer to as the tangent sheaf. \\
Once that the tangent sheaf $\mathcal{T}_\mani$ is defined, one can use the construction above to introduce its parity-reversed version $\Pi \mathcal{T}_\mani$, which is then a rank $m|n$ sheaf, locally-freely generated by $\{ \pi \partial_{\theta_1} , \ldots, \pi \partial_{\theta_{m}} | \pi \partial_{x_1}, \ldots, \pi \partial_{x_n} \}$. This sheaf $\Pi \mathcal{T}_\mani$ will play a fundamental role throughout this paper, as we shall see shortly. \\
As for the transition functions, the tangent sheaf $\mathcal{T}_\mani$ of a supermanifold $\mani$ transforms with the supertranspose of the inverse of the Jacobian of the change of coordinate
$\Phi_{ij} : \stsheaf (U_i) \lfloor_{U_i \cap U_j} \rightarrow \stsheaf (U_j) \lfloor_{U_i \cap U_j}$. Taking the parity-transpose yields the change of coordinates of $\Pi \mathcal{T}_\mani$, that so that one has for short 
\bear
M (\mathcal{T}_\mani ) = (\mathcal{J}ac (\Phi)^{-1})^{st} \qquad \qquad M (\Pi \mathcal{T}_\mani ) = ((\mathcal{J}ac (\Phi)^{-1})^{st})^{\Pi}.
\eear
Notice that this is exactly what one would find applying the chain-rule in the following form
\bear
\partial_{z_{ki}} = \sum_h \left ( \frac{\partial z_{hj}}{\partial z_{ki}} \right ) \partial_{z_{hj}}  + \sum_\ell  \left (\frac{\partial \theta_{\ell j}}{\partial z_{ki}} \right ) \partial_{\theta_{\ell j}},
\eear
that is moving the basis on the right. \\

Let us consider the example of projective superspaces $\proj {n|m}$ introduced above. In the conventions established in the previous sections, one has that the change of coordinates of $\mathcal{T}_{\proj {n|m}}$ in an intersection $U_i \cap U_j$ reads 
\begin{align}
&\partial_{z_{ji}} = z_{ij} \left ( - z_{ij} \partial_{z_{ij}} - \sum_{k\neq j, i} z_{kj} \partial_{z_{kj}} - \sum_{\kappa} \theta_{\kappa j} \partial_{\theta_{\kappa j}} \right )  \nonumber \\
&\partial_{z_{\ell i}}  = z_{ij} \partial_{z_{\ell j}} \nonumber \\
&\partial_{\theta_{\kappa i}} = z_{ij} \partial_{\theta_{\kappa j}}.
\end{align}
where here $| \partial_{z_{\ell i}} | = 0$ and $ |\partial_{\theta_{\kappa i}}| = 1.$\\
In the same intersection, the transformations for the sheaf $\Pi \mathcal{T}_{\proj {n|m}}$ instead reads: 
\begin{align}
& \pi \partial_{\theta_{\kappa i}} = z_{ij} \pi \partial_{\theta_{\kappa j}} \nonumber \\
&\pi \partial_{z_{ji}} = z_{ij} \left ( - z_{ij} \pi \partial_{z_{ij}} - \sum_{k\neq j, i} z_{kj} \pi \partial_{z_{kj}} - \sum_{\kappa} \theta_{\kappa j} \pi \partial_{\theta_{\kappa j}} \right )  \nonumber \\
&\pi \partial_{z_{\ell i}}  = z_{ij} \pi \partial_{z_{\ell j}}.
\end{align}
where now $|\pi \partial_{\theta_{\kappa i}}| = 0 $ and $|\pi \partial_{z_{\ell i}}| = 1.$\\

Let us now move to the cotangent sheaf of a supermanifold $\mani$. This is defined starting from the tangent sheaf: one puts $\mathcal{T}^{\ast}_\mani \defeq \mathcal{H}om_{\stsheaf} (\mathcal{T}_\mani, \stsheaf)$. A local basis of the cotangent sheaf, dual to $\{ \partial_{x_1}, \ldots, \partial_{x_n} | \partial_{\theta_1}, \ldots, \partial_{\theta_m} \}$ is written as usual as $\{ dx_1, \ldots, dx_n | d\theta_1, \ldots, d\theta_m \}$, where the $dx$'s are even and the $d\theta$'s are odd. The parity-reversed cotangent sheaf $\Pi \mathcal{T}^\ast_\mani$ is then a rank $m|n$ sheaf, which is locally freely-generated by $\{\pi d\theta_1, \ldots, \pi d\theta_m | \pi dx_1, \ldots, \pi dx_n \}$, where now $\pi d\theta$'s are even and the $\pi dx$'s are odd. We stress that this sheaf is usually called the \emph{sheaf of $1$-forms} and denoted with $\Omega^1_\mani$. Actually, in a completely equivalent way, one can introduce $\Pi \mathcal{T}_\mani^\ast$ as the sheaf defined by $\Pi \mathcal{T}_\mani \defeq \mathcal{H}om_{\stsheaf} (\mathcal{T}_\mani, \Pi \mathcal{O}_\mani)$, indeed
\bear
\mathcal{H}om_{\stsheaf} (\mathcal{T}_\mani, \Pi \mathcal{O}_\mani) = \mathcal{T}_\mani^\ast \otimes_{\stsheaf} \Pi \stsheaf = \Pi \mathcal{T}_\mani^\ast \otimes_{\stsheaf} \stsheaf = \Pi \mathcal{T}_\mani^\ast,
\eear
indeed $\Pi \stsheaf$ is obviously locally-free of rank $0|1$ and the functor $ - \otimes_{\stsheaf} \Pi \stsheaf$ acting on a generic sheaf of $\mathcal{O}_\mani$-modules amount exactly to the parity-change of the sheaf itself. In other words, in general, one has that if $\mathcal{E}$ is a locally-free sheaf of $\stsheaf$-modules of rank $p|q$, then one finds that $ \mathcal{E} \otimes_{\stsheaf} \Pi \stsheaf = \Pi \mathcal{E}$ is of rank $q|p$ and the transition matrices of $\mathcal{E}$ and $\Pi \mathcal{E}$ are related by a parity transposition.  \\
As for the transition functions, the cotangent sheaf transforms with the Jacobian of the change of coordinates $\Phi_{ij} : \stsheaf (U_i) \lfloor_{U_i \cap U_j} \rightarrow \stsheaf (U_j) \lfloor_{U_i \cap U_j}$, so that one finds 
\bear
M (\mathcal{T}^\ast_\mani ) = \mathcal{J}ac (\Phi) \qquad \qquad M (\Pi \mathcal{T}^\ast_\mani ) = \mathcal{J}ac (\Phi)^\Pi.
\eear
Again, this is what one would obtain by using 
\begin{align}
dz_{ki} = \sum_h \left ( \frac{\partial z_{ki}}{\partial z_{hj}} \right ) dz_{hj} + \sum_\ell \left ( \frac{\partial z_{ki}}{\partial \theta_{\ell j}}\right ) d\theta_{\ell j}.
\end{align}

Let us get back to the concrete example of the projective superspaces $\proj {n|m}$. The change of coordinates of $\mathcal{T}^\ast_{\proj {n|m}}$ reads
\begin{align}
& dz_{ji} = - \frac{dz_{ij}}{z^2_{ij}} \nonumber \\
& dz_{\ell i} = \frac{dz_{\ell j}}{z_{ij}} - \frac{z_{\ell j}}{z^2_{ij}} dz_{ji} \nonumber \\
& d\theta_{\kappa i} = \frac{d\theta_{\kappa i}}{z_{ij}} - \frac{\theta_{\kappa i}}{z^2_{ij}} dz_{ij} 
\end{align}
where $|dz_{\ell i}|= 0 $ and $|d\theta_{\kappa i }| = 1.$ The transformations of $\Pi \mathcal{T}^\ast_{\proj {n|m}}$ instead are
\begin{align}
& \pi d\theta_{\kappa i} = \frac{\pi d\theta_{\kappa j}}{z_{ij}} - \frac{\theta_{\kappa j}}{z^2_{ij}} \pi dz_{ij} \nonumber \\
&\pi dz_{ji} = - \frac{\pi dz_{ij}}{z^2_{ij}} \nonumber \\
& \pi dz_{\ell i} = \frac{\pi dz_{\ell j}}{z_{ij}} - \frac{z_{\ell j}}{z^2_{ij}} \pi dz_{ji} 
\end{align}
where now $|\pi dz_{\ell i} | = 1$ and $|\pi d\theta_{\kappa i}| = 1.$\\

Now, as should be suggested by the notation, the tangent $\mathcal{T}_\mani$ and cotangent sheaf $\mathcal{T}^\ast_\mani$, together with their parity-reversed version $\Pi \mathcal{T}_\mani$ and $\Pi \mathcal{T}^\ast_\mani$ are mutually \emph{dual}.\\
Before seeing this, though, we have to recall the following important facts of super linear algebra, that actually makes difference in computations and might lead to mistakes. First of all, let us consider a supermatrix $T$ as an \emph{even} linear transformation between certain free supercommutative modules. Writing $T$ in the block-form, one defines the supertransposition as 
\bear
T^{st} = \left ( 
\begin{array}{c|c}
A & B \\
\hline 
C & D
\end{array}
\right )^{st} \defeq \left ( 
\begin{array}{c|c}
A^t & C^t \\
\hline 
-B^t & D^t
\end{array}
\right ).
\eear
It is then immediate to see that the supertransposition has \emph{not} period 2, but 4 instead, that is $T^{st^2} \neq T$, while $T^{st^4} = T.$ Also, notice that the supertransposition does \emph{not} commute with the parity transposition, $\Pi \circ st \neq st \circ \Pi,$ but one finds instead the relations
\bear \label{reltransp}
\Pi \circ st \circ \Pi = st^2 \qquad \qquad st \circ \Pi \circ st = \Pi.
\eear
The previous discussion should warn about the issues one can encounter when dealing with the supertranspose and the parity transpose. \\
Indeed, let us now consider the case of $\mathcal{T}_\mani$ and $\mathcal{T}_\mani^\ast$. One finds
\bear
M(\mathcal{T}^\ast_\mani)^{st} \cdot M (\mathcal{T}_\mani) = \mathcal{J}ac (\Phi)^{st} \cdot (\mathcal{J}ac (\Phi)^{-1} )^{st} = (\mathcal{J}ac (\Phi)^{-1} \cdot \mathcal{J}ac (\Phi) )^{st} = id
\eear
where we have used that $(AB)^{st} = B^{st} A^{st}$. One can thus define a pairing as follows, 
\bear
\xymatrix@R=1.5pt{ 
\langle \, \cdot \,,\, \cdot \, \rangle : \, \mathcal{T}^\ast_\mani \otimes_{\stsheaf} \mathcal{T}_\mani  \ar[r] & \stsheaf \nonumber \\
\; \; \; \omega \otimes D \ar@{|->}[r] &  \langle \omega , D \rangle \defeq \omega (D)
}
\eear
for a general form $\omega$ and a vector field $D$.\\
Let us now pass to $\Pi \mathcal{T}_\mani$ and $\Pi \mathcal{T}_\mani^\ast$: one sees that it is no longer true that $M(\Pi \mathcal{T}^\ast_\mani)^{st} \cdot M (\Pi \mathcal{T}_\mani) = id$. Instead, one finds that
\begin{align}
 M(\Pi \mathcal{T}_\mani)^{st} \cdot M (\Pi \mathcal{T}^\ast_\mani )& = (\mathcal{J}ac (\Phi)^{-1})^{st \circ \Pi \circ st} \cdot \mathcal{J}ac(\Phi)^\Pi  \nonumber \\
& = (\mathcal{J}ac (\Phi)^{-1})^{\Pi} \cdot \mathcal{J}ac (\Phi)^\Pi \\
& =  ( \mathcal{J}ac (\Phi)^{-1} \cdot \mathcal{J}ac (\Phi ))^\Pi = id,
\end{align} 
where we have used the second relation in \eqref{reltransp} and that in general $(AB)^\Pi = A^{\Pi} B^{\Pi}.$ The pairing is thus written as
\bear
\xymatrix@R=1.5pt{ 
\langle \, \cdot \,,\, \cdot \, \rangle_{\Pi} : \, \Pi \mathcal{T}_\mani \otimes_{\stsheaf} \Pi \mathcal{T}^\ast_\mani  \ar[r] & \stsheaf \nonumber \\
\; \; \;  D^{\pi} \otimes \omega^{\pi} \ar@{|->}[r] &  \langle D^\pi , \omega^{\pi} \rangle_{\Pi} \defeq \omega^{\pi} (D^{\pi})
}
\eear
where $D^{\pi}$ and $\omega^\pi$ are related to $D$ and $\omega$ by a parity change. \\
Usually, one sets $\omega^{\pi} (D^{\pi}) = \omega (D)$, and it is customary to take (see for example the Appendix of \cite{Vor})
\begin{align} \label{pairing}
\left \{ 
\begin{array}{l} dz_\ell (\pi \partial_{z_\ell}) \defeq 1,  \\
d\theta_{\kappa} (\pi \partial_{\theta_\kappa})  \defeq - 1  \\
dz_\ell (\pi \partial_{\theta_\kappa} ) \defeq 0  \\
d\theta_\kappa ( \pi \partial_{z_\ell})  \defeq 0. 
\end{array}
\right.
\end{align}
Also, notice that there is \emph{no} natural pairing between $\mathcal{T}_\mani$ and $\Pi \mathcal{T}^\ast_{\mani}$ and, likewise, between $\Pi \mathcal{T}_{\mani}$ and $\mathcal{T}^\ast_\mani$: these sheaves have indeed also different rank. Nontheless, see again the Appendix in \cite{Vor} for a definition of an \virgolette odd'' pairing $\pi dz (\partial_z) = 1$ and $\pi d\theta (\partial_\theta ) = -1$, which yields an isomorphism of vector spaces and \emph{not} of vector superspaces.\\

Finally, there is one more very important natural sheaf that can be defined on a supermanifold, the so-called \emph{Berezinian sheaf}, that can be looked at as a superanalog of the canonical sheaf of an ordinary manifold, whose sections are the elements that get integrated over. The key observation is that the sections of the canonical sheaf transform as \emph{densities} under a change of local coordinates, we thus ask for a sheaf defined on the supermanifold $\mani$ whose sections transform as densities as well. This calls for finding a supergeometric analog of the notion of determinant (of an automorphism) that enters the transformations of densities such as the sections of the canonical sheaf in ordinary geometry. 
The supergeometric analog of the determinant is known as \emph{Berezianian}. Briefly, given a free $\mathbb{Z}_2$-graded module $A\defeq A^{p|q}$, the Berezianian is a supergroup homomorphisms  
\bear
\mbox{Ber} : GL (p|q; A) \longrightarrow GL(1|0; A_0)
\eear  
that agrees with the determinant when $q=0$ and it also proves to have similar properties (see \cite{Deligne} \cite{Manin} \cite{Witten}). Here $GL (p|q; A)$ are the invertible (even) automorphisms of $A$ and $A_0$ stands for the even part of $A$. \\
Given a locally-free sheaf of $\stsheaf$-modules $\mathcal{E} \leftrightarrow (\{U_i\}_{i\in I}, \{ g_{ij} (\mathcal{E})\}_{i,j \in I})$ of rank $p|q$, we thus define the Berezinian sheaf of $\mathcal{E}$ - and we denote it by $\mathcal{B}er (\mathcal{E})$ - to be the locally-free sheaf of $\mathcal{O}_\mani$-modules such that 
\bear
\mbox{rank} \, \mathcal{B}er (\mathcal{E}) = \left \{ 
\begin{array}{l} 
1|0 \qquad n + m\mbox{ even},\\
0|1 \qquad n + m \mbox{ odd}.
\end{array}
\right.
\eear 
and whose section transforms with the Berezinian $\mbox{Ber} \, g_{ij} (\mathcal{E})$ of the transition functions of $\mathcal{E}.$\\
In particular, we employ the following definition (see also \cite{Manin}): we call the Berezinian sheaf of a supermanifold $\mani$ of dimension $n|m$ the sheaf
\bear
\mathcal{B}er \mani \defeq \mathcal{B}er (\Pi \mathcal{T}_{\mani}^\ast )^\ast = \mathcal{H}om (\mathcal{B}er (\Pi \mathcal{T}_{\mani}^\ast), \mathcal{O}_\mani).
\eear
Let us see why this apparently cumbersome definition makes sense, by discussing as usual the example of the projective superspaces $\proj {n|m}.$ \\
It is well known, for example by adjunction theory, that the canonical sheaf $\mathcal{K}_{\proj n} \defeq \bigwedge^n \mathcal{T}^\ast_{\proj {n}}$ of the $n$-dimensional projective space is given by $\mathcal{K}_{\proj n} \cong \mathcal{O}_{\proj n} (-n-1)$, and indeed projective spaces are Fano manifolds, having anti-ample canonical sheaf. If we wish to obtain this result back whenever reducing to an ordinary projective space $\proj n$ from a projective superspace $\proj {n|m}$, and we wish to use the sheaf $ \Pi \mathcal{T}^{\ast}_{\proj {n|m}}$ as announced above, then we are then forced to employ the above definition. Indeed, if we are not taking the dual of the Berezinian sheaf above, we are led, because of parity reason to the \emph{wrong} relation in the case of projective spaces, getting $\mathcal{K}_{\proj n} \cong \mathcal{O}_{\proj {n}} (n+1)$ instead. 
Let us see this in some details by computing the Berezinian sheaf explicitly (see \cite{CN} for a similar computation). We start from the dual of the Euler exact sequence for projective superspaces, which is a natural generalization of the well-known Euler exact sequence for projective spaces. Upon a parity-change one gets
\bear \label{eulerform}
\xymatrix{
0 \ar[r] & \Pi \mathcal{T}_{\proj {n|m}}^\ast \ar[r] & \mathbb{C}^{n+1 |m} \otimes \Pi \mathcal{O}_{\proj {n|m}} (-1) \ar[r] & \Pi \mathcal{O}_\mani \ar[r] & 0,
}
\eear
where the sheaves $\mathcal{O}_{\proj {n|m}} (\ell)$ are again direct generalizations of the usual invertible sheaves $\mathcal{O}_{\proj n} (\ell)$, actually they are the pullback sheaves $ \pi^{-1} (\mathcal{O}_{\proj n} (\ell ) ) \otimes_{\pi^{-1} \mathcal{O}_{\proj n}} \mathcal{O}_{\proj {n|m}}$ of $\mathcal{O}_{\proj n} (\ell)$ via the projection $\pi : \proj {n|m} \rightarrow \proj n$, see again \cite{CN} for details. Taking the Berezinian of the \eqref{eulerform} yields
\bear
\mathcal{B}er (\Pi \mathcal{T}^\ast_\mani) \cong \mathcal{B}er (\mathcal{O}_{\proj {n|m}} (-1)^{\oplus m|n+1}) \cong \mathcal{O}_{\proj {n|m}} (-m+n+1),
\eear
so that, in turn
\bear
\mathcal{B}er (\Pi \mathcal{T}^\ast_\mani)^\ast \cong \mathcal{O}_{\proj {n|m}} (-m+n+1)^\ast \cong \mathcal{O}_{\proj {n|m}} (-n-1 +m).
\eear
Notice that reversing parity and tensoring the \eqref{eulerform} by $\mathcal{O}_{\proj n}$, one gets
\bear \label{eulerformsplit}
\xymatrix{
0 \ar[r] & \mathcal{T}_{\proj {n}}^\ast \oplus ({\mathcal{T}^\ast_{\proj {n|m} }}\otimes \mathcal{O}_{\proj {n}})_1 \ar[r] &  (\mathcal{O}_{\proj {n}} (-1)^{\oplus n+1})_0 \oplus (\mathcal{O}_{\proj n} (-1)^{\oplus m})_1 \ar[r] & \mathcal{O}_{\proj n} \ar[r] & 0,
}
\eear
where we have used that $(\mathcal{T}^\ast_{\proj {n|m}} \otimes \mathcal{O}_{\proj n} )_0 \cong \mathcal{T}^\ast_{\proj n}$ and whose even-reduced parts reads
\bear 
\xymatrix{
0 \ar[r] & \mathcal{T}_{\proj {n}}^\ast \ar[r] &  \mathcal{O}_{\proj {n}} (-1)^{\oplus n+1} \ar[r] & \mathcal{O}_{\proj n} \ar[r] & 0,
}\eear
as it should, so that $\mathcal{K}_{\proj n } \cong \mathcal{O}_{\proj n} (-n-1).$ The odd parts, actually yields the isomorphism $(\mathcal{T}^\ast_{\proj {n|m}} \otimes \mathcal{O}_{\proj n} )_1  \cong \mathcal{O} (-1)^{\oplus m}_1$, which can be prove to be true from very general considerations \cite{CN}.

\section{Superforms and Integral Forms Complex on a Supermanifold}

\noindent Let $\mani$ be a supermanifold of dimension $n|m$. It is possible to define the de Rham complex of differential superforms (henceforth superforms) associated to $\mani$. This is constructed starting from the sheaf $\Omega^1_\mani \defeq \Pi \mathcal{T}^\ast_\mani$, that it is locally freely-generated over an open set $U$ by
\bear
\Omega^1_\mani \lfloor_{U} \defeq \Pi \mathcal{T}^\ast_\mani \lfloor_U \cong \mathcal{O}_\mani \lfloor_U \cdot \{d\theta_1, \ldots, d\theta_m | dx_1, \ldots, dx_n \},
\eear
where we recall that $|d\theta_i |= 0$ and $|dx_j | =1$ for $i=1, \ldots, m$ and $j = 1, \ldots, n$, as seen above. There is a natural \emph{odd} differential acting as follows 
\bear
\xymatrix@R=1.5pt{ 
d: \stsheaf \ar[rr] && \Pi \mathcal{T}^\ast_\mani  \\
\qquad \quad f \ar@{|->}[rr] && df, 
}
\eear 
where $df$ is defined as
\bear
d f \defeq \sum_{i=1}^m d\theta_i \frac{\partial f}{\partial \theta_i} + \sum_{j =1}^n dx_j \frac{\partial f}{\partial x_j}, 
\eear
in agreement with \cite{Witten}, page 17, and it satisfies the $\mathbb{Z}_2$-graded Leibniz rule, as one might check,
\bear
d (f \cdot g) = df \cdot g + (-1)^{|f|} f \cdot dg,
\eear
where we have used that $|d| =1$. Importantly this differential can be lifted to an \emph{exterior derivative}
\bear
\xymatrix@R=1.5pt{ 
d^i: Sym^i \Pi \mathcal{T}^\ast_\mani  \ar[rr] && Sym^{i+1} \Pi \mathcal{T}^\ast_\mani  \\
\qquad \quad \omega \ar@{|->}[rr] && d\omega, 
}
\eear 
having the properties that $d^{i}\circ d^{i+1} = 0,$ therefore we have a complex $\Omega^{\bullet}_\mani \defeq (Sym^\bullet \Pi \mathcal{T}_\mani^\ast, d^{\bullet})$ of locally-free $\mathcal{O}_\mani$-modules as follows
\bear
\xymatrix@R=1.5pt{ 
0 \ar[r] & \stsheaf \ar[r]^d & \Pi \mathcal{T}_\mani \ar[r]^{d\;\; \; \;} & Sym^2 \Pi \mathcal{T}_\mani \ar[r]^{\; \; \; \;\; \; \;d} & \cdots \cdots \ar[r]^{d\qquad } & Sym^n \Pi \mathcal{T}_{\mani} \ar[r]^{\qquad d} & \cdots,
}
\eear
where we have dropped the index of the exterior derivative $d^i : Sym^i \Pi \mathcal{T}_\mani^\ast \rightarrow Sym^{i+1} \Pi \mathcal{T}_\mani$ for notational reasons.\\
A crucial fact should now be underlined: whilst the de Rham complex reduces - as it should - to the usual de Rham complex on a complex manifold if the the odd dimension $m$ of the supermanifold $\mani$ is zero, if $m \neq 0$ the de Rham complex on a supermanifold is \emph{not bounded from above}. In other words, there is \emph{no} notion of a top-form on a supermanifold, indeed one can actually take an arbitrary high power of the $d\theta$'s since they are commuting sections, \emph{i.e}\ $d \theta \odot \ldots \odot d\theta = d\theta^{\odot i} \neq 0 $ for any $i > 0.$\\
Let us consider for example the supermanifold $\proj {1|1}$. Then one will find that for any $i > 0$ the sheaf $Sym^i \Pi \mathcal{T}^\ast_{\proj {1|1}}$ is of rank $1|1$ and locally freely-generated over the open set $U_0$ by 
\bear
Sym^i \Pi \mathcal{T}^\ast_{\proj {1|1}} (U_0) \cong \mathcal{O}_{\proj {1|1}} (U_0) \cdot \{ d\theta^{\odot i}_{10} | dz_{10} \odot d\theta_{10}^{\odot i-1}\},
\eear 
where it is understood that $\Pi \mathcal{T}^\ast_{\proj {1|1}} (U_0) \cong \mathcal{O}_{\proj {1|1}} (U_0) \cdot \{d\theta_{10} | dz_{10} \}.$ \\

Also, notice that the Berezinian sheaf does \emph{not} appear at any place in the de Rham complex $\Omega^{\bullet}_\mani$ above, and therefore no sections of the sheaves appearing in the de Rham complex can be integrated over a supermanifold. In order to introduce a notion of top-form, suitable to define a geometric integration theory for supermanifolds, one has to resort to the notion of \emph{integral forms}. We leave to the literature \cite{Belo1, Vor, Witten} and also \cite{CCGhd, CCGSugra} a detailed discussion, here we will just sketch their main properties, in order to make the paper as self-consistent as possible. 

In particular, an \emph{integral top-form} is written locally as:
\begin{equation} 
\omega^{(n|m)}=f(x,\theta)dx_{1}\dots dx_{n}\delta(d\theta_{1})\dots
\delta(d\theta_{m})\, \label{sezioneberezin}%
\end{equation}
where $f(x,\theta) $ is a section of $\stsheaf$ and a $\mathbb{Z}_2$-graded symmetric product is understood. A single symbol $\delta (d\theta)$ is formally defined as 
\bear
\delta (d\theta ) = \int_{\mathbb{R}} e^{i d \theta  t} d t,  
\eear
where $t \in \mathbb{R}$ is an auxiliary variable, so that, referring to the expression \eqref{sezioneberezin}, one has
\bear
\delta (d\theta_1 ) \ldots \delta (d\theta_m)  \defeq \int_{\mathbb{R}^m} e^{i \sum_{i=0}^{m} d \theta_i t_i} d t_1 \wedge \ldots \wedge dt_m,  
\eear
together with their derivatives
\bear
(-i)^m \delta^{(1)} (d\theta_1) \ldots \delta^{(1)} (d\theta_m) = \int_{\mathbb{R}^m} t_1 \ldots t_m e^{i \sum_{i=0}^{m} d \theta_i t_i} d t_1 \wedge \ldots \wedge dt_m. 
\eear
Remarkably, the formal expression \eqref{sezioneberezin} transforms as a section of the Berezinian sheaf, and as such it can be integrated over.  \\

More in general, an expression involving the $dx$'s, $d\theta$'s, $\delta (d \theta)$'s and their derivatives is of the kind 
\begin{equation}
\omega^{(p|q)}=f(x,\theta)dx_{a_{1}}\dots dx_{a_{r}}d\theta_{b_{1}%
}\dots d\theta_{b_{s}}\delta^{(r_{1})}(d\theta_{c_{1}})\dots
\delta^{(r_{q})}(d\theta_{c_{q}}) \label{pseudo}%
\end{equation}
where the $\mathbb{Z}_2$-graded symmetric product between $dx, d\theta$ and $\delta$'s is understood,  
$p$ and $q$ correspond respectively to the
\textit{form} number and the \textit{picture} number, with
$0\leq q\leq m$ and $p=r+s-\sum_{i=1}^{i=q}r_{i}$ and $0 \leq r \leq n$. 
In a given monomial, the $d\theta_i$ appearing in the product are
different from those appearing in the delta's as
\bear
d\theta_i \delta (d\theta_i) =0. \label{fund}
\eear
 and
$\omega(x,\theta)$ is a set of sections of the structure sheaf, having index structure\footnote{The indices $a_{1}\dots a_{r}$ and
$b_{1}\dots b_{q}$ are anti-symmetrized, the indices $r_{1}%
\dots r_{s}$ are symmetrized because of the rules of the graded
product:
\begin{eqnarray}
dx_{a}dx_{b}  &=&-dx_{b}dx_{a}\,,~~~dx_{a}d\theta_{i}=d\theta_{i
}dx_{a}\,,~~~d\theta_{i}d\theta_{j}=d\theta_{j}d\theta_{i
}\,,\label{wedgevari}\\
\delta(d\theta_{i})\delta(d\theta_{j})  & =&-\delta(d\theta_{j
})\delta(d\theta_{i})\,,\\
~~~dx_{a}\delta(d\theta_{i})  &  =&-\delta(d\theta_{i})dx_{a}%
\,,~~~d\theta_{i}\delta(d\theta_{j})=\delta(d\theta_{j}%
)d\theta_{i}\,\,.
\end{eqnarray}} $\omega
_{\lbrack a_{1}\dots a_{r}](b_{1}\dots b_{s})[r_{1}\dots r
_{q}]}(x,\theta)$.
Also, we recall the following important rules, see for example \cite{CDGM}:
\begin{align}
& d (\delta^{(k)} (d \theta_i ) )= 0 \mbox{ for } k \geq 0,\\
& d\theta_{i}\delta^{(k)}(d\theta_{i})=-k\delta^{(k-1)}(d\theta_{i}) \mbox{ for } k>0. 
\end{align}
Notice that the meaning of the first one is that $\delta^{(k)} (d\theta) $ is $d$-closed.
With reference to the expression \eqref{pseudo}, the index $r_i$ on the delta
$\delta^{(r_i)}(d\theta_{{b_j}})$ denotes the degree of the derivative of the
delta function with respect to its argument. The total picture $q$ of
$\omega^{(p|q)}$ corresponds to the total number of delta functions and its
derivatives. The total form degree is given by $p=r+s-\sum_{i=1}^{i=q}r_{i}$ since the
derivatives act effectively as negative degree forms and the delta functions carry
zero form degree. 

In this extended scenario, we call $\omega^{(p|q)}$ a \textit{superform} if $q=0$: in this case it belongs to the honest de Rham complex $\Omega^\bullet_\mani$ we have introduced above. We call $\omega^{(p|q)}$ an \textit{integral form} if $q=m$, and we shall discuss this case in a moment; otherwise $\omega^{(p|q)}$ for $q \neq 0 , q \neq m$ is called \textit{pseudoform}.\\

Let us now take on the case of \emph{integral forms}, that is whence $q = m$. The theory of integral forms can be re-written in a manifest sheaf-theoretical formalism as to match and extend the above de Rham complex, that we will now call $\Omega^{\bullet;0}_\mani \defeq (Sym^\bullet \Pi \mathcal{T}^\ast_\mani, d^\bullet)$, as to specify the picture of the forms involved. Indeed, we claim that the integral forms fits into a new complex, we call it $\Omega^{\bullet ; m }_\mani,$ and we define it as follows
\bear
\Omega^{k ; m}_\mani \defeq ( \mathcal{B}er (\mani) \otimes Sym^{n-k} \Pi \mathcal{T}_{\mani}, d^k ), \qquad k \leq n,
\eear
where the operator $d^k : \mathcal{B}er (\mani) \otimes Sym^{n-k} \Pi \mathcal{T}_{\mani} \rightarrow \mathcal{B}er (\mani) \otimes Sym^{n-(k+1)} \Pi \mathcal{T}_{\mani} $ is induced by that defined for the de Rham complex $\Omega^{\bullet ; m}_\mani$ above, as we shall see shortly. \\
Now, it is crucial to note that the complex terminates to the Berezinian sheaf, that is $\Omega^{n;m}_\mani \defeq \mathcal{B}er (\mani)$, so that one finds 
\bear
\xymatrix{
\cdots \ar[r] & \mathcal{B}er (\mani) \otimes Sym^{n-k} \Pi \mathcal{T}_{\mani} \ar[r] & \cdots \ar[r] & \mathcal{B}er (\mani) \otimes  \Pi \mathcal{T}_{\mani} \ar[r] & \mathcal{B}er (\mani) \ar[r] & 0.
}
\eear
In order to convince the reader about the correspondence between the mathematical sheaf-theoretic formalism of the complex $\Omega^{\bullet;m}_\mani$ and the delta-function $\delta^{(k)} (d\theta)$'s formalism - which is preferred in the context of theoretical physics and string theory -, we now deal with the explicit and simple example of $\proj {1|1}$.\\
We aim to match the modules involving the delta's for a fixed total form degree $k$, with the sheaves 
\bear
\Omega^{k;1}_{\proj {1|1}} \defeq \mathcal{B}er (\proj {1|1}) \otimes_{\proj {1|1}} Sym^{1-k} \Pi \mathcal{T}_{\proj {1|1}}
\eear 
for any $k \leq 1$ appearing in the complex $\Omega^{\bullet, m}_\mani$, by comparing their transition functions in the only intersection $U_0 \cap U_1$ of $\proj {1|1}$. \\
In order to do that we start dealing with the delta's formalism, as we first need the transformation properties the integral forms on projective superspaces. Generalizing the result of \cite{CDGM}, and following the convention set above, on a general projective superspace $\proj {n|m}$ one finds
\begin{align} \label{transdelta}
\delta^{(0)} (d\theta_{\ell i}) & = \delta^{(0)} \left ( \frac{d\theta_{\ell j}}{z_{ij}} - \frac{\theta_{\ell j}}{z^2_{ij}} dz_{ij} \right ) = z_{ij} \delta^{(0)} \left ({d\theta_{\ell j}} - \frac{\theta_{\ell j}}{z_{ij}} dz_{ij} \right ) \nonumber \\
& = z_{ij} \delta^{(0)} \left ( d\theta_{\ell j} \right ) - \theta_{\ell j} dz_{ij} \delta^{(1)} (d\theta_{\ell j}) \nonumber \\
& = z_{ij} \delta^{(0)} \left ( d\theta_{\ell j} \right ) + \theta_{\ell j} \delta^{(1)} (d\theta_{\ell j})dz_{ij},
\end{align}
where we recall that $|\delta^{i} (d\theta)| = 1$ and we have Taylor expanded (the increment has been put to the left) around the $d\theta$.  \\
Generalizing this formula, for higher-derivatives one finds
\begin{align} \label{transdeltader}
\delta^{(k)} (d\theta_{\ell i}) = z^{k+1}_{ij} \delta^{(k)} \left ( d\theta_{\ell j} \right ) + z^{k}_{ij} \theta_{\ell j} \delta^{(k+1)} (d\theta_{\ell j})dz_{ij}.
\end{align}
Now, we have that in the delta's formalism the modules are locally generated by expressions of the kind 
\bear
\Omega^{k;1}_{\proj {1|1}}(U_0) \cong \mathcal{O}_{\proj {1|1}}(U_0) \cdot \{dz_{10} \delta^{(1-k)} (d\theta_{10}) | \delta^{(-k)} (d\theta_{10})  \} \qquad k < 1,
\eear
and $\Omega^{1;1}_{\proj {1|1}} (U_0) \cong \mathcal{O}_{\proj {1|1}}(U_0) \cdot \{ dz_{10} \delta^{(0)} (d\theta_{10})\}$. Using the transformation properties in the \eqref{transdelta} and \eqref{transdeltader} adapted for $\proj {1|1},$ one finds the following transition matrix
\bear
M (\Omega^{1;1}_{\proj {1|1}}) = \left ( - \frac{1}{z} \right ) \qquad 
M (\Omega^{k;1}_{\proj {1|1}}) = \left ( \begin{array}{c|c}
- z^{-k} & 0 \\
\hline 
 \theta z^{-k} & z^{1-k} 
\end{array}
\right ) \mbox{ for } k \leq 1,
\eear
where we have dropped for convenience the indices referring to the only intersection $U_0 \cap U_1$ on $\proj {1|1}.$\\

Let us now look at the sheaf-theoretic formalism. First of all, we have that, $\mathcal{B}er (\proj {1|1}) \cong \mathcal{O}_{\proj {1|1}} (-1)$. Moreover, if $\mathcal{B}er (\proj {1|1})$ is locally-generated over $U_0$ by $\mathcal{D} [dz_{10} |d \theta_{10}]$ (see for example \cite{Witten} for this notation), calculating explicitly the transition function of this rank $1|0$ locally-free sheaf, one has 
\bear
\mathcal{D} [dz_{10} |d \theta_{10}] = \left (- \frac{1}{z_{01}} \right) \mathcal{D} [dz_{01} |d \theta_{01}],
\eear
thus matching $M(\Omega^{1;1}_{\proj {1|1}})$ above, as expected.\\
Also, locally, for $k < 1$, one finds that 
\bear
\mathcal{B}er (\proj {1|1}) \otimes Sym^{1-k} \Pi \mathcal{T}_{\proj {1|1}} (U_0) = \mathcal{O}_{\proj {1|1}} (U_0) \cdot \mathcal{D} [dz_{10} |d \theta_{10}]  \otimes  \left \{ \pi \partial_{\theta_{10}}^{\odot 1-k} \Big |  \pi \partial_{z_{10}} \odot \pi \partial_{\theta_{10}}^{\odot - k} \right \}
\eear 
and, using the transformation rules introduced above for the sheaves of the kind $\Pi \mathcal{T}_{\proj {n|m}}$ specialized to the case of $\proj {1|1}$, one finds that the transformation matrix reads
\bear
M (\mathcal{B}er (\proj {1|1}) \otimes Sym^{1-k} \Pi \mathcal{T}_{\proj {1|1}}) = - \frac{1}{z} \otimes \left ( \begin{array}{c|c}
z^{1-k} & 0 \\
\hline 
- \theta z^{1 -k} & - z^{2-k} 
\end{array}
\right ) = \left ( \begin{array}{c|c}
- z^{-k} & 0 \\
\hline 
 \theta z^{-k} & z^{1-k} 
\end{array}
\right ),
\eear
thus matching the remaining ones for $k<1$.\\
We therefore has the following correspondence, realizing the actual isomorphism between the sheaf-theoretic and the delta's formalism:
\begin{align}
& \mathcal{D}[dz|d\theta] \otimes \pi \partial_{\theta}^{\odot 1-k} \longleftrightarrow dz \delta^{(1-k)} (d\theta),\\
& \mathcal{D}[dz|d\theta] \otimes \pi \partial_z \odot \pi \partial_{\theta}^{\odot -k} \longleftrightarrow \delta^{(-k)} (d\theta),
\end{align}
for $k <1$, together with $\mathcal{D}[dz|d\theta] \leftrightarrow dz \delta^{(0)} (d\theta),$ that are sections of the Berezinian sheaf $\Omega^{1;1}_{\proj {1|1}} \cong \mathcal{B}er (\proj {1|1}).$
\\

Now that the correspondence is achieved at the level of the sheaves, we still have to deal with the coboundary operator $d$ of the complex $\Omega^{\bullet ; m}_{\proj {1|1}}$. First, we recall that looking at $\proj {1|1}$ one has a differential, acting locally as
\bear
\xymatrix@R=1.5pt{ 
d_U : \mathcal{O}_{\proj {1|1}}(U) \ar[r]  & \Pi \mathcal{T}_{\proj {1|1}}^\ast (U)  \\
F \ar@{|->}[r] & d_UF \defeq dz \,\partial_z F + d\theta \,\partial_\theta F,
}
\eear
that lifts to an exterior differential for the de Rham complex $\Omega^{\bullet ; 0}_{\proj {1|1}} $, as observed above. 
Now we recall that in \eqref{pairing} we have set a pairing on the local generator of $\Pi \mathcal{T}_{\proj {1|1}}$ and $\Pi \mathcal{T}^\ast_{\proj {1|1}}$ as follows:
\begin{align} \label{usualpairing}
\left \{ 
\begin{array}{l} d\theta (\pi \partial_\theta) \equiv \langle \pi \partial_\theta , d\theta \rangle \defeq - 1 \nonumber \\
dz(\pi \partial_z) \equiv \langle \pi \partial_z , dz \rangle \defeq 1, \nonumber \\
dz (\pi \partial_\theta) \equiv \langle \pi \partial_\theta, dz \rangle \defeq 0 \nonumber \\
d\theta ( \pi \partial_z) \equiv \langle \pi \partial_z , d \theta \rangle \defeq 0. 
\end{array}
\right.
\end{align}
These relations will be used to extend the differential to the integral forms, in order to construct a honest complex.\\
Indeed, let $s^{(k)} \in \mathcal{B}er (\proj {1|1}) \otimes Sym^{1-k} \Pi \mathcal{T}_{\proj {1|1}} = \Omega^{k;1}_{\proj {1|1}}$ be a generic integral forms. Then, in a certain chart
\bear
s^{(k)} = \mathcal{D}[dz|d\theta] \otimes \left ( F  \cdot (\pi \partial_\theta^{\odot 1 - k}) + G \cdot (\pi \partial_z \odot \pi \partial^{-k}_\theta ) \right  )
\eear
for some $F,G \in \mathcal{O}_{\proj {1|1}}.$ \\
We thus put
\bear
d s^{(k)} \defeq \mathcal{D}[dz|d\theta] \otimes \left ( dF  \cdot (\pi \partial_\theta^{\odot 1 - k}) + dG \cdot (\pi \partial_z \odot \pi \partial^{-k}_\theta ) \right  ),
\eear
so that one finds 
\begin{align}
d s^{(k)} = \mathcal{D}[dz|d\theta] \otimes \left ( (dz \, \partial_z F + d\theta\, \partial_\theta F) \cdot (\pi \partial_\theta^{\odot 1 - k}) + (dz \,\partial_z G + d\theta \,\partial_\theta G) \cdot (\pi \partial_z \odot \pi \partial^{\odot -k}_\theta ) \right ),
\end{align}
and upon using the pairing defined above, one has 
\bear
ds^{(k)} = \mathcal{D}[dz|d\theta] \otimes \left (  -|1-k| \partial_\theta F \cdot (\pi \partial_\theta^{\odot - k} ) + (-1)^{|G|} \partial_z G \cdot (\pi \partial_\theta^{\odot -k }) - |k| \partial_\theta G \cdot (\pi \partial_z \odot \pi \partial_\theta^{\odot -k-1})\right ) \nonumber \\. 
\eear
Note that this defines a section in $ \mathcal{B}er (\proj {1|1}) \otimes Sym^{-k} \Pi \mathcal{T}_{\proj {1|1}} = \Omega^{k+1;1}_{\proj {1|1}},$ as it should. \\
Applying again $d$ yields
\bear
d (ds^{(k)}) = \mathcal{D}[dz|d\theta] \otimes \left ( (-1)^{|G|} d\theta \, \partial_\theta \partial_z G \cdot (\pi \partial_\theta^{\odot -k }) - |k| dz \, \partial_z\partial_\theta G \cdot (\pi \partial_z \odot \pi \partial_\theta^{\odot -k-1}) \right ).
\eear
Using again the pairings above, one gets:
\begin{align}
d (ds^{(k)}) & = \mathcal{D}[dz|d\theta] \otimes \left (- |k| (-1)^{|G|} \partial_\theta \partial_z G \cdot (\pi \partial_\theta^{\odot -k-1}) - |k| (-1)^{|G|+1} \partial_z \partial_\theta G \cdot (\pi \partial_\theta^{\odot -k-1} ) \right ) \nonumber \\ 
& = |k|(-1)^{|G|} \mathcal{D}[dz|d\theta] \otimes \left (- \partial_z \partial_\theta G  + \partial_z \partial_\theta G \right ) \cdot ( \pi \partial_\theta^{\odot -k-1} )=0,
\end{align}
as $[\partial_z, \partial_\theta] = 0.$\\
This shows that $d\circ d = 0$, so it can be promoted to a coboundary operator for the complex of integral forms $( \Omega^{k ; 1}_{\proj {1|1}} = \mathcal{B}er (\proj {1|1}) \otimes Sym^{1-k} \Pi \mathcal{T}_{\proj {1|1}}, d^k)$, with  
\bear
\xymatrix@R=1.5pt{ 
d^k : \mathcal{B}er (\proj {1|1}) \otimes Sym^{1-k} \Pi \mathcal{T}_{\proj {1|1}} \ar[r] & \mathcal{B}er (\proj {1|1}) \otimes Sym^{1-(k+1)} \Pi \mathcal{T}_{\proj {1|1}},
}
\eear 
so that 
\begin{align}
\xymatrix@R=1.5pt{
\ldots \ar[r]^{d \quad \qquad \qquad } & \mathcal{B}er (\proj {1|1}) \otimes Sym^{1-k} \Pi \mathcal{T}_{\proj {1|1}}  \ar[r]^{\quad \qquad \qquad d} & \ldots \ar[r]^{ d   \quad \; \qquad } & \mathcal{B}er (\proj {1|1}) \otimes \Pi \mathcal{T}_{\proj {1|1}} \ar[r]^{\qquad \; d} & \mathcal{B}er (\proj{1|1}) \ar[r] & 0 \  }
\end{align}
This simple example can be generalized to any supermanifold $\mani$, as to yield its integral forms complex.\\

Moreover, the fundamental relations characterizing the delta's, \emph{i.e.}\ equation \eqref{fund} and following, can be recovered in a more geometric fashion using the sheaf-theoretic formalism developed above. In particular, one finds that the basic relation $d\theta \delta^{(0)} (d\theta) = 0$ can be recover using the pairing defined above. Indeed, we recall that one has $ \delta^{(0)} (d\theta)= \mathcal{D} (dz|d\theta) \otimes \pi \partial_{z},$ so that one finds 
\bear
d\theta \delta^{(0)} (d\theta)  = 0 \quad  \longleftrightarrow \quad \mathcal{D} (dz|d\theta) \otimes \langle \pi \partial_z , d \theta \rangle = 0.
\eear
Even more, for higher-derivatives of the delta's, one has $dz d\theta \delta^{(1)}(d\theta)  = - \delta^{(0)}(d\theta) dz$. Recalling that $dz  \delta^{(1)}(d\theta) =  \mathcal{D} (dz|d\theta) \otimes \pi \partial_{\theta} $, one finds again via the pairing 
\bear
dz d\theta \delta^{(1)}(d\theta)  = - dz  \delta^{(0)}(d\theta) \quad \longleftrightarrow \quad \mathcal{D} (dz|d\theta) \otimes \langle \pi \partial_{\theta} , d\theta \rangle = -\mathcal{D} (dz|d\theta)
\eear
where we recall that $\mathcal{D} (dz|d\theta)$ is indeed a generating section of the Berezinian sheaf, corresponding to $\delta^{(0)} (d\theta) dz$ in the integral forms delta's notation.\\

Actually, as the attentive reader might have already noticed, it is fair to say that the whole construction of integral forms above can be obtained from first principles starting from the de Rham complex $\Omega^{\bullet;0}_\mani$, upon using some homological algebra. This construction is completely natural and spare us from cumbersome choices. Leaving the details to future works, we just observe that, indeed, for a completely generic supermanifold $\mani$ of dimension $n|m$, the locally-free sheaves making up the complex of integral forms $\Omega^{\bullet; m}_\mani$ can be obtained from those appearing in the de Rham complex, $\Omega^{\bullet ; 0}_\mani = Sym^\bullet \Pi \mathcal{T}^\ast_\mani$, simply by applying the functor $h_{\mathcal{B}er (\mani)} \defeq \mathcal{H}om_{\stsheaf} (-, \mathcal{B}er (\mani))$ to them. In other words, one has
\bear
\xymatrix{
\Omega^{k;0}_\mani \ar@{|->}[rr]^{h_{\mathcal{B}er (\mani)} \qquad \qquad \qquad} && \mathcal{H}om_{\mathcal{O}_\mani} (\Omega^{k;0}_\mani, \mathcal{B}er (\mani)) =  \Omega^{n-k; m}_\mani 
}\eear
for $k \geq 0.$ Also, notice that this functor is \emph{contravariant}: this means that if one has a morphism of sheaves $\phi : \Omega^{k, 0}_\mani \rightarrow \Omega^{k+1;0}_\mani$, then applying the functor $h_{\mathcal{B}er (\mani)} $ to $\phi$ yields a morphism $h_{\mathcal{B}er (\mani)} (\phi) \defeq \mathcal{H}om_{\mathcal{O}_\mani} (\phi, \mathcal{B}er (\mani)) : \Omega^{n-k-1;m}_\mani \rightarrow \Omega^{n-k}_\mani$. In particular, recalling that by definition functors preserve composition, one has $h_{\mathcal{B}er (\mani)} (\phi \circ \psi ) = h_{\mathcal{B}er (\mani)} (\phi ) \circ h_{\mathcal{B}er (\mani)} ( \psi )$. Choosing $\psi = d^k : \Omega^{k;0}_\mani \rightarrow \Omega^{k+1;0}_\mani$ and $\phi = d^{k+1}: \Omega^{k+1;0} \rightarrow \Omega^{k+2;0}_\mani$, one gets that $0 = h_{\mathcal{B}er (\mani)} (d^{k+1} \circ d^{k} ) = h_{\mathcal{B}er (\mani)} (d^{k+1} ) \circ h_{\mathcal{B}er (\mani)} ( d^k )$, which thus makes $(h_{\mathcal{B}er (\mani)} (\Omega^{\bullet;0}_\mani), h_{\mathcal{B}er (\mani)} (d^\bullet) )$ into a (cochain) complex, actually the integral forms complex. The homological features of this construction and their implications will be elucidated in a forthcoming paper.\\

Before we go on, a remark is in order, though. While on the one hand we have already seen that superforms and integral forms are well-behaved and that they can be given a structure of complexes of locally-free \emph{finitely generated} sheaves of $\mathcal{O}_\mani$-modules, this is no longer true for \emph{pseudoforms} - that is for middle-dimensional picture $0 < p<m $. Indeed, it can be seen that, for a fixed form number, pseudoforms are locally arranged in modules that are \emph{not} finitely-generated, and therefore they cannot be described globally as locally-free sheaves of finite rank, such as for superforms and integral forms. \\
Let us see this by means of the easiest possible example, that of $\proj {1|2}$, which has already been discussed in \cite{1DCY}. \\
Using the delta's formalism, one sees that those modules having picture number equal to 1 are generated over the open set $U_0$ by expressions of the kind
\begin{align} \label{middleforms}
\Omega^{k ; 1}_{\proj {1|2}} (U_0) = \mathcal{O}_{\proj {1|2}} (U_0) \cdot & \Big \{ \delta^{(\ell+1)}(d \theta_{10} ) dz_{10} d\theta_{20}^{k+\ell +1}, \delta^{(\ell+1)}(d \theta_{20} ) dz_{10} d\theta_{10}^{k+\ell +1} \,\big | \nonumber \\
&  \big |\, \delta^{(\ell)}(d \theta_{10} ) d\theta_{20}^{k+\ell } , \delta^{(\ell)}(d \theta_{20} ) d\theta_{10}^{k+\ell}  \Big \}  \qquad \ell \in \mathbb{N} \cup \{ 0\}.
\end{align}
This example suggests that these expressions can be are arranged in \emph{quasi-coherent sheaves} of $\mathcal{O}_{\proj {1|2}}$-modules - and as such they might have an infinite \v{C}ech cohomology, see \cite{1DCY} -. By they way a more careful description of this particular quasi-coherent sheaves would be necessary in order to get a complete mathematically satisfying picture of the zoo of forms on a supermanifold.

\section{Negative Degree Superforms and their Complex}

\noindent We have seen that integral forms allow to define complexes of forms carrying a negative degree. Moreover, as explained in the section \ref{LHSPCO}, in the framework of Large Hilbert Space, also ordinary superforms carrying a negative degree make their appearance: we will call these new special superforms: \emph{inverse superforms}. \\     
The fundamental observation is once again that, given a supermanifold $\mani$, the local sections $d\theta$'s of the sheaf $\Pi \mathcal{T}_{\mani}^\ast$ are \emph{even}, and therefore they can be formally \virgolette inverted''. Here, we describe this phenomenology making use of our usual driving example of $\proj {1|1}$. Notice that since we are interested into the algebraic-geometric properties of these special superforms, we will describe them as sections of certain sheaves, and we will make no distinction between $1/d\theta$ and $\mbox{p.v.} (1/d\theta)$, as their transformation properties coincide. \\
Let us consider the sheaf $\Omega^{1;0}_{\proj {1|1}} = \Pi \mathcal{T}_{\proj {1|1}}^\ast$. As seen, over the open set $U_0$ of $\proj {1|1}$, $\Pi \mathcal{T}_{\proj {1|1}}^\ast$ is locally-freely generated by $\{ d\theta_{10} | dz_{10}\}.$ We would like to get a form of degree equal to $-1$, by \virgolette dividing'' by $d \theta_{10}$, so we consider formal expressions like 
\bear
\Omega^{-1; 1}_{\proj {1|1}}(U_0) = \mathcal{O}_{\proj {1|1}} (U_0) \cdot \left \{ \frac{1}{d\theta_{10}} \Big | \frac{dz_{10}}{d\theta_{10}^{ 2}} \right \}\,. 
\eear
Notice that, as for the sheaves of pseudoforms, only whenever the supermanifold $\mani$ is of odd dimension equal to $1$, the sheaves $\Omega_\mani^{k;p}$ of any degree and picture number are coherent, actually locally-free sheaves of $\mathcal{O}_\mani$-modules. Indeed, let us look again at the case of $\proj {1|2}$. One would find that the expressions with degree $-1$ can be generated by 
\bear \label{p12meno}
\Omega^{-1;0}_{\proj {1|2}} (U_0) = \mathcal{O}_{\proj {1|2}} (U_0) \cdot \Bigg \{ \frac{d\theta^{\ell}_{10}}{d\theta_{20}^{ \ell+1}}, \frac{d\theta^{\ell}_{20}}{d\theta_{10}^{ \ell +1}} \Bigg | dz_{10}\frac{d\theta^{\ell}_{10}}{d\theta_{20}^{ \ell + 2}}, dz_{10}\frac{d\theta^{\ell}_{20}}{d\theta_{10}^{ \ell + 2}} \Bigg \} \qquad \ell \in \mathbb{N} \cup \{0\}\,.
\eear
One thus sees that these expressions are likely to make up a quasi-coherent sheaf, but not certainly a locally-free sheaf of finite rank, an issue that again makes the theory more difficult. \\
Restricting ourself to the case having a single odd dimension, as in the previous section, we now aim to give this formal setting a sheaf-theoretic dignity. One can see that, in general, for $\proj {1|1}$
\bear
\Omega^{-k;0 }_{\proj {1|1}} (U_0) \cong \mathcal{O}_{\proj {1|1}} (U_0) \cdot \Bigg \{ \frac{1}{d\theta_{10}^{ k}} \Bigg | \frac{dz_{10}}{d\theta_{10}^{ k+1}} \Bigg \} \qquad k > 0.
\eear
In particular, for $k=1$, the transformations read
\begin{align}
&  \frac{1}{d\theta_{1 0}} = z_{01} \frac{1}{d\theta_{1 1}} + \theta_{1 1} \frac{dz_{01}}{d\theta_{1 1}^2}, \\
& \frac{dz_{10}}{d\theta^2_{1 0} } = - \frac{dz_{01}}{d\theta^2_{1 1} }, 
\end{align}
such that, generalizing to a generic $k>0$, one gets a transformation matrix of the form:
\bear
M(\Omega^{-k;0}_{\proj {1|1}}) = \left ( \begin{array}{c|c}
 z^k & k z^{k-1} \theta \\
\hline 
 0 &  - z^{k-1}
\end{array}
\right ).
\eear 
Now, one can see that the case $k=1$ corresponds to the transition functions of the sheaf $\Pi \mathcal{B}er (\proj {1|1}) \otimes \Pi \mathcal{T}_{\proj {1|1}}, $ indeed:
\bear
M(\Pi \mathcal{B}er (\proj {1|1}) \otimes \Pi \mathcal{T}_{\proj {1|1}}) = \left ( \begin{array}{c|c}
z & \theta \\
\hline 
 0 & -1 
\end{array}
\right ),
\eear
as one can readily verify, settling the first case. Notice that in this case $ \Pi \mathcal{B}er (\proj {1|1}) \otimes \Pi \mathcal{T}_{\proj {1|1}} \cong  \mathcal{B}er (\proj {1|1}) \otimes  \mathcal{T}_{\proj {1|1}}.$\\
More in general one finds that the correspondence we are looking for is 
\bear
\Omega^{-k;0}_{\proj {1|1}} \cong \Pi \mathcal{B}er (\proj {1|1}) \otimes Sym^k \Pi \mathcal{T}_{\proj {1|1}} \qquad k>0,
\eear 
where we observe that the functors $\Pi$ and $Sym^k$ do not commute (for $k > 1$), so that $\Pi \circ Sym^k \neq Sym^k \circ \Pi.$ Actually, the transition matrix of the sheaf $\Pi \mathcal{B}er (\proj {1|1}) \otimes Sym^k \Pi \mathcal{T}_{\proj {1|1}}$ is given by 
\bear
M(\Pi \mathcal{B}er (\proj {1|1}) \otimes Sym^k \Pi \mathcal{T}_{\proj {1|1}}) = \left ( \begin{array}{c|c}
 z^k & z^{k-1} \theta \\
\hline 
 0 &  - z^{k-1}
\end{array}
\right ),
\eear
 but clearly the numerical factor $k$ in the upper-right entry of the matrix above can be recovered by a constant change of basis, actually a constant scaling of the generators - that does not modify the class in the cohomology set of the transition functions and, hence it does not change the sheaf we have identified. \\
In particular, choosing 
\bear 
A = \left ( \begin{array}{c|c}
\sqrt{k} & 0 \\
\hline 
0 & 1/\sqrt{k} 
\end{array}
\right )
\eear
does the job, as
\bear
A  \left ( \begin{array}{c|c}
z^k & \theta z^{k-1} \\
\hline 
0 & - z^{k-1}
\end{array}
\right )  A^{-1} = \left ( \begin{array}{c|c}
 z^k & k z^{k-1} \theta \\
\hline 
 0 &  - z^{k-1}
\end{array}
\right ).
\eear
Also, notice incidentally that $\det A = 1$.
To conclude, the correspondence goes as follows: 
\begin{align}
&\frac{1}{d\theta^k} \quad \longleftrightarrow \quad \sqrt{k} (\pi \mathcal{D}[dz|d\theta]) \otimes (\pi \partial_{z} \odot \pi \partial_\theta^{\odot k-1} ) \\
&\frac{dz}{d\theta^{k+1}} \quad \longleftrightarrow \quad \frac{1}{\sqrt{k}} (\pi \mathcal{D}[dz|d\theta]) \otimes (\pi \partial_\theta^{\odot k}),
\end{align}
thus giving the sheaf-theoretic identification:
\bear
\Omega^{-k;0}_{\proj {1|1}} \cong \Pi \mathcal{B}er (\proj {1|1}) \otimes Sym^{k} \Pi \mathcal{T}_{\proj {1|1}} \qquad  k>0.
\eear
The case $k = 0 $ deserves a special attention, indeed the sheaf $\Omega^{0;0}_{\proj {1|1}}$ - that formerly corresponded to the structure sheaf $\mathcal{O}_{\proj {1|1}}$ - gets modified to 
\bear \label{case0}
\Omega^{0;0}_{\proj {1|1}} \cong \mathcal{O}_{\proj {1|1}} \oplus \Pi \mathcal{B}er (\proj {1|1}),
\eear
as one has to take into account also an element which is locally of the form $\frac{dz}{d\theta}$, and therefore it is a superform of degree zero. It is straighforward to verity that such an element transforms as a section of parity-changed Berezinian sheaf $\Pi \mathcal{B}er (\proj {1|1})$, that is $\frac{dz}{d\theta} \equiv \pi \mathcal{D}[dz | d\theta]$. We will see in a moment why the structure sheaf gets extended in this way. \\

As we got this far, a crucial fact that has to be noted is that the sheaves making up the inverse superforms of a certain fixed degree, corresponds to certain sheaves of integral forms of certain fixed degree, but having \emph{opposite parity}. In particular, it can be seen that that for $k\leq 0$ one has
\bear
\underbrace{\Omega^{k;1}_{\proj {1|1}} = \mathcal{B}er (\proj {1|1}) \otimes Sym^{1-k} \Pi \mathcal{T}_{\proj {1|1}}}_{\mbox{\tiny{Integral Forms}}} \quad \longleftrightarrow \quad \underbrace{\Omega^{k-1;0}_{\proj {1|1}} =\Pi \mathcal{B}er (\proj {1|1}) \otimes Sym^{1-k} \Pi \mathcal{T}_{\proj {1|1}}}_{\mbox{\tiny{Inverse Forms}}}.
\eear 
It follows that, altogether, one gets the following diagram:
\bear \label{diagp1}
\xymatrix{
\ldots \ar[r] &  \Pi \mathcal{B}er (\proj {1|1}) \otimes \Pi \mathcal{T}_{\proj {1|1}} \ar[r] & \Pi \mathcal{B}er (\proj {1|1})\oplus \mathcal{O}_{\proj {1|1}} \ar[r] & \ar[r] \Pi \mathcal{T}^\ast_{\proj {1|1}} \ar[r] & \ldots   \\
\ldots \ar[r] &  \mathcal{B}er (\proj {1|1}) \otimes Sym^2 \Pi \mathcal{T}_{\proj {1|1}}  \ar[ul]_{\Theta} \ar[r] & \mathcal{B}er (\proj {1|1}) \otimes \Pi \mathcal{T}_{\proj {1|1}}  \ar[ul]_{\Theta} \ar[r] & \mathcal{B}er (\proj {1|1}) \ar[ul]_{\Theta} \ar[r] & 0
}
\eear
Here, the map $\Theta \defeq \Theta (\iota_{\partial_\theta})$ introduced in sec 2, acts as the parity changing functor $\Pi$, and as such it is an \emph{isomorphism up to the parity}. Note that the first map $\Theta : \mathcal{B}er (\proj {1|1}) \hookrightarrow \Pi \mathcal{B}er (\proj {1|1}) \oplus \mathcal{O}_{\proj {1|1}}$ is an immersion (again, up to parity) instead, acting as $s \mapsto (\pi s , 0).$ In this way, using the geometric language developed, 
\bear 
\xymatrix@R=1.5pt{ 
\Theta \equiv \Pi : \Omega^{k; 1} \equiv \mathcal{B}er (\proj {1|1}) \otimes Sym^{1-k} \Pi \mathcal{T}_{\proj {1|1}} \ar[r] & \Omega^{k-1;0} \equiv \Pi \mathcal{B}er (\proj {1|1}) \otimes Sym^{1-k} \Pi \mathcal{T}_{\proj {1|1}}   \\
s \ar@{|->}[r] & \pi s
}
\eear
for $k< 1$. Explicitly, using the bases chosen above, one finds:
\bear
\xymatrix@R=1.5pt{
 \mathcal{D}[dz|d\theta] \otimes \pi \partial_\theta^{\odot -k} \ar[rr]^{\Theta} &&  \frac{1}{\sqrt{k}} (\pi \mathcal{D}[dz|d\theta]) \otimes (\pi \partial_\theta^{\odot - k}) \\
 \mathcal{D}[dz|d\theta] \otimes ( \pi \partial_\theta^{\odot 1-k} \odot \pi \partial_z )  \ar[rr]^{\Theta} &&  \sqrt{k} (\pi \mathcal{D}[dz|d\theta]) \otimes (\pi \partial_\theta^{\odot 1-k} \odot \pi \partial_{z} ).
 }
\eear
The above discussion leads to the following picture: one sees that allowing for inverse superforms, that is expressions of the kind $\frac{1}{d\theta}$, corresponds to enlarge the usual complex of superforms on the supermanifold - in our case $\proj {1|1}$ - by a \virgolette copy'' of the complex of integral forms having opposite parity, shifted to the left by a number of steps equal to the odd dimension of the supermanifold. 
Also, observe that the exterior differentials defined for the integral forms and the superforms as in the previous section can be used to make this sequence of sheaves into an actual complex, so that with abuse of notation, we write:
\begin{align}
\xymatrix{
\cdots \ar[r] & \Pi \mathcal{B}er(\proj {1|1}) \otimes \Pi \mathcal{T}_{\proj {1|1}} \ar[r]^d & \Pi \mathcal{B}er (\proj {1|1}) \oplus \mathcal{O}_{\proj {1|1}} \ar[r]^{\qquad \; \; d} & \Pi \mathcal{T}_{\proj {1|1}}^\ast \ar[r]^{d\quad } & Sym^2 \Pi \mathcal{T}^\ast_{\proj {1|1}} \ar[r] & \cdots  
} 
\end{align}

Notice that, even if the transition functions will be more complicated, everything we have said above can be repeated almost identically also in the case of a generic supermanifold $\mani$ of dimension $n|1$. Keeping on with the case of projective spaces, considering $\proj {n|1}$, one has that for the integral forms
\bear
\Omega^{k;1}_{\proj {n|1}} \; \longleftrightarrow \; \mathcal{B}er (\proj {n|1}) \otimes Sym^{n-k} \Pi \mathcal{T}_{\proj {n|1}}
\eear
for $k\leq n$. And likewise for the inverse superforms,
\bear
\Omega^{-k;0}_{\proj {n|1}} \; \longleftrightarrow \; \Pi \mathcal{B}er (\proj {n|1}) \otimes Sym^{n-1+k} \Pi \mathcal{T}_{\proj {n|1}}
\eear
for $k > 0.$ \\
Also, we have an odd morphism $\Theta : \Omega^{k;1}_{\proj {n|1}} \rightarrow \Omega^{k-1;0}_{\proj {n|1}}$ that for $k\leq 0$ is an \emph{isomorphism up to the parity}. In the sheaf-theoretic formalism it reads
\bear 
\xymatrix@R=1.5pt{ 
\Theta_{k\geq 1}\equiv \Pi : \mathcal{B}er (\proj {n|1}) \otimes Sym^{n-k} \Pi \mathcal{T}_{\proj {n|1}} \ar[r] & \Pi \mathcal{B}er (\proj {n|1}) \otimes Sym^{n-k} \Pi \mathcal{T}_{\proj {n|1}}   \\
s \ar@{|->}[r] & \pi s
}
\eear
for $k \leq 0$ and it is indeed just a parity-inversion.\\
Differently, the morphism $\Theta  : \Omega^{k;1}_{\proj {n|1}} \rightarrow \Omega^{k-1;0}_{\proj {n|1}}$ for $0 < k \leq n$ is just an \emph{injective} morphism, as made clear by the sheaf-theoretic formalism, indeed:
\bear
\xymatrix@R=1.5pt{ 
\Theta_{k > 0} : \mathcal{B}er (\proj {n|1}) \otimes Sym^{n-k} \Pi \mathcal{T}_{\proj {n|1}} \ar[r] & \left ( \Pi \mathcal{B}er (\proj {n|1}) \otimes Sym^{n-k} \Pi \mathcal{T}_{\proj {n|1}} \right ) \oplus Sym^{k-1} \Pi \mathcal{T}^\ast_{\proj {n|1}} \\
s \ar@{|->}[r] & ( \pi s, 0)
}
\eear  
This gives the following realization for picture number $p=0$ of this extended de Rham complex, where the superforms have been supplemented by the inverse forms as well:
\bear \label{enlarged}
\Omega_{\proj {n|1}}^{k;0} \cong \left \{ 
\begin{array}{ll}  
Sym^{k-1} \Pi \mathcal{T}^{\ast}_{\proj {n|1}} & k > n \\
\left ( \Pi \mathcal{B}er (\proj {n|1}) \otimes Sym^{n-k} \Pi \mathcal{T}_{\proj {n|1}} \right ) \oplus Sym^{k-1} \Pi \mathcal{T}^\ast_{\proj {n|1}} &  0 < k \leq n \\
\Pi \mathcal{B}er (\proj {n|1}) \otimes Sym^{n-k} \Pi \mathcal{T}_{\proj {n|1}} & k\leq 0.
\end{array}
\right.
\eear
Getting back to the example of $\proj {1|1}$, we have that the sheaf-theoretic representation of the Large Hilbert Space related to the supermanifold $\proj {1|1}$, we denote it by $\mathcal{LHS}_{\proj {1|1}}$, is given by 
\bear \label{p1lhs}
\mathcal{LHS}_{\proj {1|1}} = \left \{ 
\begin{array}{ll} 
Sym^k \Pi \mathcal{T}^{\ast}_{\proj {1|1}} & k > 0\\
\left ( \mathcal{O}_{\proj {1|1}} \oplus \Pi \mathcal{B}er (\proj {1|1}) \right ) \oplus \mathcal{B}er (\proj {1|1}) & k = 0\\
 \left ( \Pi \mathcal{B}er (\proj {1|1}) \otimes Sym^{|k|} \Pi \mathcal{T}_{\proj {1|1}} \right ) \oplus \left ( \mathcal{B}er (\proj {1|1}) \otimes Sym^{|k|} \Pi \mathcal{T}_{\proj {1|1}}   \right )  & k < 0. 
\end{array}
\right.
\eear
In agreement with what have been said in the section 2, we stress the \virgolette duplication'' arising in this enlarged context: clearly, besides the case $k>0$ where only usual positive-degree forms appears, the other cases displays all of the elements belonging to the superforms complex - enlarged by the inverse superforms - together with all of the elements belonging to the integral forms complex. The Large Hilbert Space $\mathcal{LHS}_{\proj {1|1}}$ therefore is the copy of two identical sheaves, $\Omega^{k;0}_{\proj {1|1}} \oplus \Omega_{\proj {1|1}}^{k+1;1} \cong \Omega^{k;0}_{\proj {1|1}} \oplus \Pi \Omega_{\proj {1|1}}^{k;0}$, having opposite parity in the case $k < 0$. The case $k=0$ is somehow \virgolette critical'', as in the extended superforms complex the structure sheaf is supplemented by the Berezinian sheaf coming from the integral form complex lifted via $\Theta \equiv \Pi $, as shown in \eqref{case0}, and in the case $k>0$ there are the usual superforms only, because there are no integral forms to be lifted.\\

Actually, before we conclude this section, it is fair to say that in String Field Theory the Large Hilbert Space related to a certain supermanifold is structured - better than just as sheaf of modules as above -, as a sheaf of algebras, with a formal notion of multiplication between superforms on the one hand and integral forms on the other hand. This, in turn, is to be viewed as a structure inherited by the extended superforms complex $\Omega^{k;0}_{\mani}$, for $k \in \mathbb{Z}$, that is given as well a formal notion of multiplication between superforms, mimicking the exterior (or better, supersymmetric) product $\Omega^{k_1;0}_\mani \wedge \Omega^{k_2;0}_\mani \rightarrow \Omega^{k_1 +k_2;0}_\mani $, so that on the local generators, one formally puts
\begin{align} \label{ciccio}
& d\theta^{k_1} \wedge \dfrac{1}{d\theta^{k_2}} = d\theta^{k_1-k_2}, \nonumber \qquad \quad d\theta^{k_1} \wedge \dfrac{dz}{d\theta^{k_2}} = dz d\theta^{k_1-k_2}, \nonumber \\
& dz d\theta^{k_1} \wedge \dfrac{1}{d\theta^{k_2}} = dz d\theta^{k_1-k_2}, \nonumber \qquad \quad dz d\theta^{k_1} \wedge \dfrac{dz}{d\theta^{k_2}} = 0,   
\end{align}
where, clearly, $d\theta^{k_1-k_2} = 1/d\theta^{|k_1 - k_2|}$ if $k_1< k_2$. Recovering these relations and endowing the extended superforms complex with a honest algebra structure making the formal relations above rigorous, is not straightforward as one might expect given the modules description provided above, and we leave this to a future paper. Nonetheless, this is possibly where the $A_\infty$-algebra structure appearing in string field theory enters the description, thus providing an appealing relation between the underlying supergeometry on which the theory relies on, and $A_\infty$-algebras, which would be something that is worth to be carefully investigated \cite{Erler1, Erler2, Erler3}.

\subsection{Large Hilbert Space and \v{C}ech Cohomology} It is not hard to provide the \v{C}ech cohomology of the Large Hilbert Space in the example of $\proj {1|1}$ we have dealt with so far. Indeed, by super Serre duality (see the last section of the present paper for an explanation), it is fully determined by the \v{C}ech cohomology of the case $k\geq0$ in the \eqref{p1lhs} only - which in turn amount to compute the usual \v{C}ech cohomology of forms on a supermanifold. \\
We will do this in two ways. First, we treat the explicit example of $\proj {1|1}$, exploiting two facts: 
$\proj {1|1}$ is a projected supermanifold, and its reduced manifold is the Riemann sphere. Indeed, since $\proj {1|1}$ is a projected supermanifold, every locally-free sheaf of $\mathcal{O}_{\proj {1|1}}$-modules $\mathcal{E}_{\mathcal{O}_{\proj {1|1}}}$, such as $Sym^k \Pi \mathcal{T}_{\proj {1|1}}$, is also a locally-free sheaf of $\mathcal{O}_{\proj {1}}$-modules $\mathcal{E}_{\mathcal{O}_{\proj 1}}$, and therefore, by virtue of the Grothendieck Theorem for locally-free sheaves on $\proj {1}$, it splits into a sum of invertible sheaves of the kind $\mathcal{O}_{\proj {1}} (k)$ for $k\in \mathbb{Z}$, see \cite{Har}, that is,
\bear
\mathcal{E}_{\mathcal{O}_{\proj 1}} \cong \bigoplus_{i=1}^n \mathcal{O}_{\proj {1}} (k_i), \qquad \quad k_i \in \mathbb{Z}
\eear
where we have forget about the parity and where $n = p+q$ is the rank of $\mathcal{E}_{\mathcal{O}_{\proj 1}}$ if $\mathcal{E}_{\mathcal{O}_{\proj {1|1}}}$ is of rank $p|q$. In particular, we have that $Sym^k \Pi \mathcal{T}^\ast_{\proj {1|1}}$ is locally freely-generated over $\mathcal{O}_{\proj 1}$ by $\{ d\theta^{\odot k}, \theta d\theta^{\odot k-1} \odot dz | dz\odot  d\theta^{\odot k-1}, \theta d\theta^{\odot k} \}, $ so that the matrix of the transition functions can be written using the usual rules as
\bear
[M (Sym^k \Pi \mathcal{T}^\ast_{\proj {1|1}})] = \left ( 
\begin{array}{cc|cc}
1/z^{k} & - 1/z^{k+1} & 0 & 0 \\
0 & - 1/z^{k+2} & 0 & 0 \\
\hline 
0 & 0 & -1/k^{k+1} & 0 \\
0 & 0 & 0 & - 1/z^{k+1} 
\end{array}
\right ). 
\eear 
By simple (allowed) rows and columns operations this matrix can be brought to diagonal form  
\bear
[M (Sym^k \Pi \mathcal{T}^\ast_{\proj {1|1}})] = \left ( 
\begin{array}{cc|cc}
1/z^{k+1} & 0 & 0 & 0 \\
0 & - 1/z^{k+1} & 0 & 0 \\
\hline 
0 & 0 & -1/k^{k+1} & 0 \\
0 & 0 & 0 & - 1/z^{k+1} 
\end{array}
\right ), 
\eear
so that one can read from this expression the factorization into invertible sheaves over $\proj 1$ of the sheaf $Sym^k \Pi \mathcal{T}^{\ast}_{\proj {1|1}}$ for $k\geq 1$:
\bear
Sym^k \Pi \mathcal{T}^{\ast}_{\proj {1|1}} \cong \mathcal{O}_{\proj 1} (-k-1)^{\oplus 2} \oplus \Pi \mathcal{O}_{\proj 1} (-k-1)^{\oplus 2}.
\eear
The cohomology is easily computed: 
\bear
H^{0} (Sym^k \Pi \mathcal{T}^{\ast}_{\proj {1|1}} ) \cong 0, \qquad H^{1} (Sym^k \Pi \mathcal{T}^{\ast}_{\proj {1|1}} ) \cong \mathbb{C}^{2k|2k}
\eear
where again $k\geq 1.$
By super Serre duality, one gets that for $k \geq 1$
\bear
H^{0} (\mathcal{B}er (\proj {1|1}) \otimes Sym^k \Pi \mathcal{T}_{\proj {1|1}} )^\ast \cong \mathbb{C}^{2k|2k}, \qquad H^1 (\mathcal{B}er (\proj {1|1}) \otimes Sym^k \Pi \mathcal{T}_{\proj {1|1}}) \cong 0.
\eear
Also, $H^0 (\mathcal{O}_{\proj {1|1}}) \cong \mathbb{C}^{1|0} \cong H^1 (\mathcal{B}er (\proj {1|1}))$ and $H^1 (\mathcal{O}_{\proj {1|1}}) \cong 0 \cong H^0 (\mathcal{B}er (\proj {1|1}))$, so that one can see the cohomology of the extended de Rham complex and, in turns, of the Large Hilbert Space:
\begin{align}
&H^{0} (\mathcal{LHS}_{\proj {1|1}} )\cong \left \{ 
\begin{array}{ll} 
0  & k > 0\\
\mathbb{C}^{1|0}\oplus 0 & k = 0\\
\mathbb{C}^{2|k| \, |\, |2|k|} \oplus \mathbb{C}^{2|k| \, |\, |2|k|} & k < 0, 
\end{array}
\right. \\
&H^{1} (\mathcal{LHS}_{\proj {1|1}} )\cong \left \{ 
\begin{array}{ll} 
\mathbb{C}^{2k \, |\, 2k} & k > 0\\
0 \oplus \mathbb{C}^{0|1} & k = 0\\
0  & k < 0. 
\end{array}
\right.
\end{align}

The second methods we introduce is more general and holds true for any projective superspace of the kind $\proj {n|m}$. Indeed, let us start from the super analog of the (dual of the) Euler exact sequence for the cotangent sheaf \cite{CN}, \cite{ManinNC}:
\bear \label{eulercot}
\xymatrix{
0 \ar[r] & \mathcal{T}^{\ast}_{\proj {n|m}}  \ar[r] & \mathcal{O}_{\proj {n|m }} (-1)^{\oplus n+1|m} \ar[r] &  \mathcal{O}_{\proj {n|m}} \ar[r] & 0.
}
\eear
Its parity inverted version - which is the one we are interested into - reads
\bear
\xymatrix{
0 \ar[r] & \Pi \mathcal{T}^{\ast}_{\proj {n|m}}  \ar[r] & \mathcal{O}_{\proj {n|m }} (-1)^{\oplus m |n+1} \ar[r] &  \Pi \mathcal{O}_{\proj {n|m}} \ar[r] & 0.
}
\eear
Now, since we are interested into the cohomology of $Sym^k \Pi \mathcal{T}^\ast_{\proj {n|m}}$, we have to consider its $k$-symmetric power. We observe that since $\Pi \mathcal{O}_{\proj {n|m}}$ is of rank $0|1$, we will have that $Sym^k \Pi \mathcal{O}_{\proj {n|m}} \cong 0$ if $k \geq 2$ and that, by definition of exact sequence of sheaves, locally 
\bear 
\mathcal{O}_{\proj {n|m }} (-1)^{\oplus m |n+1} \stackrel{loc}{\cong } \Pi \mathcal{T}^{\ast}_{\proj {n|m}} \oplus \Pi \mathcal{O}_{\proj {n|m}}. 
\eear
It follows that, taking the $k$-symmetric power, one gets
\begin{align}
(Sym^k \mathcal{O}_{\proj {n|m }})^{\oplus m |n+1} (-k) & \stackrel{loc}{\cong } Sym^k \Pi \mathcal{T}^\ast_{\proj {n|m}} \oplus Sym^{k-1}
\Pi \mathcal{T}^\ast_{\proj {n|m}} \otimes \Pi \mathcal{O}_{\proj {n|m}}\nonumber \\  
& \cong  Sym^k \Pi \mathcal{T}^\ast_{\proj {n|m}} \oplus \Pi Sym^{k-1} \Pi \mathcal{T}^\ast_{\proj {n|m}},  
\end{align}
since $Sym^k \Pi \mathcal{O}_{\proj {n|m}} \cong 0$ for any $k\geq 2$, as observed above. This implies that the $k$-symmetric power of the exact sequence \eqref{eulercot} reads
\bear \label{sympi}
\xymatrix{
0 \ar[r] & Sym^k \Pi \mathcal{T}^{\ast}_{\proj {n|m}}  \ar[r] & ( Sym^k \mathcal{O}_{\proj {n|m }}^{\oplus m |n+1}) (-k) \ar[r] &  \Pi Sym^{k-1} \Pi \mathcal{T}^\ast_{\proj {n|m}} \ar[r] & 0,
}
\eear
which tells that the cohomology of the sheaf $Sym^k \Pi \mathcal{T}^{\ast}_{\proj {n|m}} $ can be computed recursively. Notice incidentally that the short exact sequence \eqref{sympi} is actually completely general, it is deduced from completely general considerations, and it can be used to compute the sheaf cohomology of forms for any projective supermanifolds by restriction, once the embedding $\iota: \mani \hookrightarrow \proj {n|m}$ is given.\\
Let us check that the result for $\proj {1|1}$ matches with what we have found above by means of Grothendieck Theorem. Noticing that, in general, for $\proj {1|1}$ we have that $(Sym^k \mathcal{O}_{\proj {1|1}}^{\oplus 1|2}) (-k) \cong \mathcal{O}_{\proj {1|1}} (-k)^{\oplus {2|2} }$, the short exact sequence \eqref{sympi} reads 
\bear \label{sympiA}
\xymatrix{
0 \ar[r] & Sym^k \Pi \mathcal{T}^{\ast}_{\proj {1|1}}  \ar[r] & \mathcal{O}_{\proj {1|1 }} (-k)^{\oplus 2 |2} \ar[r] &  \Pi Sym^{k-1} \Pi \mathcal{T}^\ast_{\proj {1|1}} \ar[r] & 0,
}
\eear
for $k \geq 1.$ One finds that the only non trivial terms in the long exact cohomology sequence are those in the following short exact sequence of vector superspaces:
\bear \label{sympiB}
\xymatrix{
0 \ar[r] & H^1 (Sym^k \Pi \mathcal{T}^{\ast}_{\proj {1|1}})   \ar[r] & H^1 (\mathcal{O}_{\proj {1|1 }} (-k))\otimes \mathbb{C}^{\oplus 2 |2} \ar[r] &  H^1 (\Pi Sym^{k-1} \Pi \mathcal{T}^\ast_{\proj {1|1}} )\ar[r] & 0.
}
\eear
It is very easy to see that 
\bear
H^1 (\mathcal{O}_{\proj {1|1 }} (-k))\otimes \mathbb{C}^{\oplus 2 |2} \cong \mathbb{C}^{\oplus k-1|k } \otimes \mathbb{C}^{2|2} \cong \mathbb{C}^{\oplus 4k-2 |4k -2},
\eear
thus by recursion one sees that $ H^1 (\Pi Sym^{k-1} \Pi \mathcal{T}^\ast_{\proj {1|1}} ) \cong \mathbb{C}^{\oplus 2k-2 |2k-2}$, and in turns 
\bear
H^1 (Sym^k \Pi \mathcal{T}^{\ast}_{\proj {1|1}}) \cong \mathbb{C}^{\oplus 2k|2k}, 
\eear
just as above.

\subsection{Large Hilbert Space and Calabi-Yau Supermanifolds} In this brief subsection we keep on looking at the extended forms complexes and its related Large Hilbert Space, by taking on a slightly different-flavored example, that of Calabi-Yau supermanifolds, \emph{i.e.}\ supermanifolds having trivial Berezinian sheaf. \\
In doing so we will deal with possibly the easiest example of Calabi-Yau supermanifold in genus $0$, which is given by the so-called $\Pi$-projective line $\proj {1}_{\Pi}$. As recently  explained in \cite{PiGeo} in much greater generality for a generic $\Pi$-projective space $\proj n_{\Pi}$, the $\Pi$-projective line can be looked at as the classifying space of the $1|1$-dimensional $\Pi$-symmetric sub superspaces of $\mathbb{C}^{2|2}$, that is all those sub superspaces $S$ that are stable under the action of a morphism $p_\Pi : S \rightarrow \Pi S$, with $p_{\Pi}^2 = id$, which is a representation of the parity changing functor in the category of vector superspaces. Clearly, given a vector superspace $\mathbb{C}^{n|n} = \mathbb{C}^{n} \oplus \Pi \mathbb{C}^{n}$ we can choose a basis of even elements as follows $\mathbb{C}^n = \mbox{Span} \{ e_1, \ldots, e_n \} $. Starting from these elements, we can obtain a basis for the whole $\mathbb{C}^{n|n}$ by putting $\mathbb{C}^{n|n} = \mbox{Span} \{ e_1, \ldots, e_n \,| \, p_{\Pi} e_1, \ldots, p_{\Pi}e_n \}$. Here, the action of $p_{\Pi} : \mathbb{C}^{n|n} \rightarrow \Pi \mathbb{C}^{n|n} \cong \mathbb{C}^{n|n}$ exchanges the generators of $\mathbb{C}^{n}$ with those of $\Pi \mathbb{C}^{n}.$ This picture has a very simple consequence: if we are given a vector superspace $V^{n|n}$ together with a basis $\{ e_1, \ldots, e_n \,| \, p_{\Pi} e_1, \ldots, p_{\Pi}e_n \}$, then a sub vector superspace of $V^{n|n}$ is $\Pi$-symmetric if and only if for every element $v = \sum_{i=1}^n z^i e_i + \theta^i p_{\Pi} e_i$ it also contains $v_{\Pi} = \sum_{i = 1}^n (- \theta^i e_i  + z^i p_\Pi e_i )$.\\
This last point of view is useful to realize $\proj {1}_{\Pi}$ as a closed supermanifold inside a super Grassmannian, namely $\mathbb{G} (1|1 ; \mathbb{C}^{2|2})$, see \cite{Manin} or \cite{NojaPhD}: this is covered by two affine superspaces, each isomorphic to $\mathbb{C}^{1|1}$, having coordinates in the \emph{super big-cells} notation given by
\begin{align}
\mathcal{Z}_{\mathcal{U}_0} \defeq 
\left ( \begin{array}{ccc||ccc} 
1 & & z_0 & 0 & &  \theta_0 \\
\hline \hline
0 & & - \theta_0  & 1 & & z_0
\end{array}
\right ) \qquad \qquad
\mathcal{Z}_{\mathcal{U}_1} \defeq
\left ( \begin{array}{ccc||ccc} 
z_1 & & 1 & \theta_1 & &  0 \\
\hline \hline
-\theta_1 & & 0  & z_1 & & 1 
\end{array}
\right ).
\end{align}
The transition functions in the intersections of the charts can be found by (allowed) rows and column operation, yielding \cite{PiGeo}: 
\begin{align}
& \left ( \begin{array}{cc||cc} 
1 &  z_0 & 0 &   \theta_0 \\
\hline \hline
0 & - \theta_0  & 1  & z_0
\end{array}
\right )  \stackrel{R_0 / z_0, R_1 /z_0}{\longrightarrow} 
\left ( \begin{array}{cc||cc} 
1/z_0 & 1 & 0 &   \theta_0/z_0 \\
\hline \hline
0 &  - \theta_0/z_0  & 1/z_0 & 1
\end{array}
\right ) \nonumber \\
& \left ( \begin{array}{cc||cc} 
1/z_0 & 1 & 0  &  \theta_0/z_0 \\
\hline \hline
0 & - \theta_0/z_0  & 1/z_0 & 1
\end{array}
\right ) \stackrel{R_0 - \theta_0/z_0 R_1 }{\longrightarrow} \left ( \begin{array}{cc||cc} 
1/z_0 & 1 & - \theta_0/z_0^2 & 0\\
\hline \hline
0 & - \theta_0/z_0  & 1/z_0 & 1
\end{array}
\right ) \nonumber \\
& \left ( \begin{array}{cc||cc} 
1/z_0  & 1 & - \theta_0/z_0^2 & 0\\
\hline \hline
0 & - \theta_0/z_0  & 1/z_0 & 1
\end{array}
\right ) \stackrel{R_1 + \theta_0/z_0 R_0 }{\longrightarrow} 
\left ( \begin{array}{cc||cc} 
1/z_0 & 1 & - \theta_0/z_0^2 & 0\\
\hline \hline
\theta_0/z_0^2 & 0  & 1/z_0 & 1
\end{array}
\right ) \nonumber.
\end{align}
The transition functions characterizing the structure sheaf $\mathcal{O}_{\proj {1}_\Pi}$ of the $\Pi$-projective line can be read from the above expression and one gets  
\begin{equation} 
z_1 = \frac{1}{z_0}, \qquad \qquad \xi_1 = - \frac{\theta_1}{z_0^2}.
\end{equation}
It follows that the $\Pi$-projective line $\proj {1|1}_\Pi $ is the $1|1$-dimensional supermanifold characterised by the pair $(\proj {1}, \mathcal{O}_{\proj 1} (-2) \cong {\Omega^{1}_{\proj 1}}),$ and it is easy to get that the $\mathcal{B}er (\proj 1_{\Pi}) \cong \mathcal{O}_{\proj 1_\Pi},$ that is $\proj 1_{\Pi}$ is a Calabi-Yau supermanifold. This has a particular nice consequence, that is the sheaves of integral forms simplifies to: 
\bear
\Omega^{k;1}_{\proj {1}_\Pi} = Sym^{k-1} \Pi \mathcal{T}_{\proj {1}_{\Pi}}.
\eear
Notice that this is a peculiarity which is true - up to parity - for all of the Calabi-Yau supermanifold: for example, the well-known \emph{super twistor-space} $\proj {3|4} (= \mathbb{CP}^{3|4})$, which is a Calabi-Yau supermanifold because $\mathcal{B}er (\proj {3|4}) \cong \Pi \mathcal{O}_{\proj {3|4}}$, has sheaves of integral forms given by $\Omega^{k;4}_{\proj {3|4}} = \Pi Sym^{k-3} \Pi \mathcal{T}_{\proj {3|4}}$.\\
Looking at the generators of the sheaves of integral forms for the $\Pi$-projective line, one gets the following correspondence between generators for $\Omega^{k;1}_{\proj {1}_\Pi} = Sym^k \Pi \mathcal{T}_{\proj {1}_\Pi}:$ 
\begin{align} \label{corrp1}
& \pi \partial_{\theta}^{\odot 1-k} \longleftrightarrow dz \delta^{(1-k)} (d\theta),\nonumber \\
& \pi \partial_z \odot \pi \partial_{\theta}^{\odot -k} \longleftrightarrow \delta^{(-k)} (d\theta),
\end{align}
for $k \leq 0$, together with $\mathcal{B}er (\proj {1}_\Pi) \owns \mathcal{D}[dz|d\theta] = s \in \mathcal{O}_{\proj 1_\Pi}$, as $\proj 1_\Pi$ is a Calabi-Yau supermanifold.\\ 
Likewise, the same arguments as above apply to recover the inverse forms characterizing the extended form complex: 
\bear \label{enlargedP1}
\Omega_{\proj 1_\Pi}^{k;0} \cong \left \{ 
\begin{array}{ll}  
Sym^{k} \Pi \mathcal{T}^{\ast}_{\proj {1}_\Pi} & k > 0 \\
\mathcal{O}_{\proj 1_{\Pi}} \oplus \Pi \mathcal{O}_{\proj {1}_\Pi} &  k = 0 \\
\Pi Sym^{|k|} \Pi \mathcal{T}_{\proj {1}_\Pi} & k < 0.
\end{array}
\right.
\eear
Switching to a more comfortable notation, notice in particular that for the sheaves $\Omega^{-k;0}_{\proj 1_\Pi} = \Pi Sym^{k} \Pi \mathcal{T}_{\proj {1}_\Pi}$ when $k >0$ the correspondence between the generators is 
\begin{align} \label{correspondenceforms1}
& \pi ( \pi \partial_z \odot \pi \partial_{\theta}^{\odot k -1}) \longleftrightarrow \dfrac{1}{d\theta^{k}},\\
& \pi ( \pi \partial_{\theta}^{\odot k}) \longleftrightarrow \dfrac{dz}{d\theta^{k+1}}. \label{correspondenceforms2}
\end{align}
Likewise, the Large Hilbert Space of the Calabi-Yau supermanifold $\proj 1_\Pi$ is given by
\bear
\mathcal{LHS}_{\proj {1}_\Pi} = \left \{ 
\begin{array}{ll} 
Sym^k \Pi \mathcal{T}^{\ast}_{\proj {1}_\Pi} & k > 0\\
\left ( \mathcal{O}_{\proj {1}_\Pi} \oplus \Pi \mathcal{O}_{\proj {1}_\Pi} \right ) \oplus \mathcal{O}_{\proj {1}_\Pi} & k = 0\\
 \left ( \Pi Sym^{|k|} \Pi \mathcal{T}_{\proj {1}_\Pi} \right ) \oplus \left (  Sym^{|k|} \Pi \mathcal{T}_{\proj {1}_\Pi}   \right )  & k < 0. 
\end{array}
\right.
\eear
It is again easy to compute the \v{C}ech cohomology of the Large Hilbert Space using Grothendieck Theorem. Looking at the splitting as locally-free $\mathcal{O}_{\proj 1}$-modules of the sheaf $Sym^k \Pi \mathcal{T}^\ast_{\proj 1_\Pi}$ over $\proj 1_\Pi$ for $k\geq 1$, one finds the following matrix of transition functions
\bear
[M (Sym^k \Pi \mathcal{T}^\ast_{\proj {1|1}})] = \left ( 
\begin{array}{cc|cc}
(-1)^k/z^{2k} &  (-1)^{k+1}/z^{2k+1} & 0 & 0 \\
0 &  (-1)^{k+1}/z^{2k+2} & 0 & 0 \\
\hline 
0 & 0 & (-1)^{k}/k^{2k+1} & 0 \\
0 & 0 & 0 & (- 1)^{k+1}/z^{2k+2} 
\end{array}
\right ), 
\eear
that can be diagonalized by rows and columns operations to 
\bear
[M (Sym^k \Pi \mathcal{T}^\ast_{\proj {1|1}})] = \left ( 
\begin{array}{cc|cc}
(-1)^k/z^{2k+1} &  0 & 0 & 0 \\
0 &  (-1)^{k+1}/z^{2k+1} & 0 & 0 \\
\hline 
0 & 0 & (-1)^{k}/k^{2k} & 0 \\
0 & 0 & 0 & (- 1)^{k+1}/z^{2k+2} 
\end{array}
\right ), 
\eear
so that one has the following decomposition
\bear
Sym^k \Pi \mathcal{T}^\ast_{\proj 1_\Pi} \cong \mathcal{O}_{\proj 1} (-2k-1)^{\oplus 2} \oplus \Pi \left ( \mathcal{O}_{\proj 1} (-2k) \oplus \mathcal{O}_{\proj 1} (-2k-2)\right ). 
\eear
The cohomology is therefore given by
\bear
H^{0} (Sym^k \Pi \mathcal{T}^{\ast}_{\proj {1}_\Pi} ) \cong 0, \qquad H^{1} (Sym^k \Pi \mathcal{T}^{\ast}_{\proj {1}_\Pi} ) \cong \mathbb{C}^{4k|4k},
\eear
that in turn yields the following cohomology for the Large Hilbert Space:
\bear
H^0 (\mathcal{LHS}_{\proj 1_\Pi} ) = \left \{ 
\begin{array}{ll} 
0 & k > 0\\
\left ( \mathbb{C}^{1|0} \oplus \mathbb{C}^{0|1} \right ) \oplus \mathbb{C}^{1|0} & k = 0\\
 \mathbb{C}^{4k|4k} \oplus \mathbb{C}^{4k|4k}  & k < 0, 
\end{array}
\right.
\eear
\bear
H^1 (\mathcal{LHS}_{\proj 1_\Pi} ) = \left \{ 
\begin{array}{ll} 
\mathbb{C}^{4k|4k} & k > 0\\
\left ( \mathbb{C}^{0|1} \oplus \mathbb{C}^{1|0} \right ) \oplus \mathbb{C}^{0|1} & k = 0\\
 0  & k < 0.
\end{array}
\right.
\eear

\section{Superforms and Pseudo-forms for Higher Odd Dimensions}

\noindent Let us now look at what happens in the case one deals with a supermanifold $\mani$ having odd dimension greater than $1$. To keep the discussion as concrete as possible we keep on using the example of $\proj {1|2}$. We have already seen in the expression \eqref{p12meno} at the beginning of the previous section, that as soon as one allows superforms of negative degree, the sheaves $ \Omega_{\proj {1|2}}^{k;0} $ for $k \in \mathbb{Z}$, making up the \virgolette extended'' de Rham complex are no longer locally-free sheaves of a certain (finite) rank, but infinitely generated quasi-coherent sheaves instead.\\
Including also \emph{pseudo-forms} of middle dimensional picture number - which are similarly arranged into infinitely generated quasi-coherent sheaves $\Omega^{k; 1}_{\proj {1|2}} $, for $k \in \mathbb{Z}$, as shown in \eqref{middleforms} -, the whole picture goes as follows: 
\bear \label{diag}
\xymatrix@C=1.7cm{
\cdots \ar[r] & \Omega^{-2;0}_{\proj {1|2}} \ar[r] & \Omega^{-1;0}_{\proj{1|2}} \ar[r] & \Omega^{0;0}_{\proj {1|2}} \ar[r] & \Omega^{1;0}_{\proj {1|2}} \ar[r] & \cdots  \\
\ldots \ar[r] & \Omega^{-2;1}_{\proj{1|2}} \ar[r] & \ar@/^/[ul]^{\Theta_1}  \ar@/_/[ul]_{\Theta_2} \Omega^{-1;1}_{\proj {1|2}} \ar[r] & \Omega^{0;1}_{\proj {1|2}} \ar@/^/[ul]^{\Theta_1}  \ar@/_/[ul]_{\Theta_2} \ar[r] & \Omega^{1;1}_{\proj{1|2}} \ar[r] & \cdots \\
\cdots \ar[r] & \Omega^{-2;2}_{\proj {1|2}} \ar[r] & \Omega^{-1;2}_{\proj {1|2}} \ar[r] &\ar@/^/[ul]^{\Theta_1}  \ar@/_/[ul]_{\Theta_2}  \Omega^{0;2}_{\proj {1|2}} \ar[r] & \ar@/^/[ul]^{\Theta_1}  \ar@/_/[ul]_{\Theta_2} \Omega^{1;2}_{\proj {1|2}} \ar[r] & 0. \\
}
\eear
We recall that the bottom line, corresponding to integral forms, is a complex of locally-free sheaves of $\mathcal{O}_{\proj{1|2}}$-modules, and in particular, one has, for $ k \in \mathbb{Z}$, $k \leq 1,$
\bear
\Omega^{k;2}_{\proj {1|2}} \defeq \mathcal{B}er (\proj {1|2}) \otimes Sym^{|k-1|} \Pi \mathcal{T}_{\proj {1|2}} \cong \Pi Sym^{|k-1|} \Pi \mathcal{T}_{\proj {1|2}},
\eear
since $\proj {1|2}$ is a Calabi-Yau supermanifold, in that $\mathcal{B}er (\proj {1|2}) \cong \Pi \mathcal{O}_{\proj {1|2}}.$\\
As explained in the previous section, one has that $\Theta_{i}$ for $i=1,2$ is a sheaf morphism as follows
\bear
\Theta_i \defeq \Theta (\iota_{\theta_i}) : \Omega^{k;p}_{\proj {1|2}} \longrightarrow \Omega^{k-1;p-1}_{\proj {1|2}}.
\eear
Using the formal expressions involving the delta's and the inverse of superforms, they act in a certain chart as $\delta^{(i)} (d\theta_i ) \stackrel{\Theta_i}{\longmapsto} {1}/{d\theta_i^{i+1}}$, so that for example, starting from the bottom line of integral forms and acting with $\Theta_1$ one has
\begin{align}
\xymatrix@R=0.1cm{
\Theta_1 : \Omega^{1;2}_{\proj {1|2}} \ar[rr] && \Omega^{0;1}_{\proj {1|2}} \\
dz \delta^{(0)} (d\theta_1) \delta^{(0)} (d\theta_2) \ar@{|->}[rr] && \frac{dz}{d\theta_1} \delta^{(0)} (d \theta_2) 
}\\
\xymatrix@R=0.1cm{
\Theta_1 : \Omega^{0;2}_{\proj {1|2}} \ar[rr] && \Omega^{-1;1}_{\proj {1|2}} \\
{ \left ( \begin{array}{l} 
\delta^{(0)} (d\theta_1) \delta^{(0)} (d\theta_2) \\
dz \delta^{(1)} (d\theta_1) \delta^{(0)} (d\theta_2) \\
dz \delta^{(1)} (d\theta_1) \delta^{(0)} (d\theta_2) 
\end{array}
\right ) }  \ar@{|->}[rr] && 
{ \left ( \begin{array}{l} 
\frac{\delta^{(0)} (d\theta_2)}{d\theta_1} \\
\frac{dz\delta^{(0)} (d\theta_2) }{d\theta_1^2}   \\
\frac{dz\delta^{(1)} (d\theta_2)}{d\theta_1}  
\end{array}
\right )
}}
\end{align}
Studying carefully the transition functions of the above expressions one can see that, once again, as in the case of $\proj {1|1}$, the morphisms $\Theta_i$ are nothing but a change of parity, that is $\Theta_i \equiv \Pi.$ \\
Less formally, recalling that $\Gamma \mathcal{B}er (\proj {1|2}) \owns \mathcal{D}[dz| d\theta_1, d\theta_2] \equiv dz \delta^{(0)} (d\theta_1) \delta^{(0)} (d\theta_2) $ and that $\Omega^{0;2}_{\proj {1|2}} \cong \mathcal{T}_{\proj {1|2}}$, so that $\partial_z \equiv \delta^{(0)} (d\theta_1) \delta^{(0)} (d\theta_2) $ and $\partial_{\theta_1} \equiv dz \delta^{(1)} (d\theta_1) \delta^{(0)} (d\theta_2) $ and $\partial_{\theta_2} \equiv dz \delta^{(0)} (d\theta_1) \delta^{(1)} (d\theta_2)$, one finds that 
\begin{align}
\xymatrix@R=0.1cm{
\Theta_1 \equiv \Pi : \mathcal{B}er (\proj {1|2}) \ar[rr] && \Omega^{0;1}_{\proj {1|2}} \\
\mathcal{D}[dz| d\theta_1, d\theta_2]  \ar@{|->}[rr] && \pi \mathcal{D}[dz| d\theta_1, d\theta_2]
}\\
\xymatrix@R=0.1cm{
\Theta_1 \equiv \Pi  : \mathcal{T}_{\proj {1|2}} \ar[rr] && \Omega^{-1;1}_{\proj {1|2}} \\
{ \left ( \begin{array}{l} 
\partial_z \\
\partial_{\theta_1} \\
\partial_{\theta_2} 
\end{array}
\right ) }  \ar@{|->}[rr] && 
{ \left ( \begin{array}{l} 
\pi \partial_z \\
\pi \partial_{\theta_1} \\
\pi \partial_{\theta_2} 
\end{array}
\right ),
}}
\end{align}
and the same applies to $\Theta_2 \equiv \Pi.$ Notice that codomains of the map $\Theta_1$ are still denoted as $\Omega^{0;1}_{\proj {1|2}}$ and $\Omega^{-1;1}_{\proj {1|2}}$ respectively since they are not identified yet as known sheaf, such as the domains instead. In particular, turning back to the formal language, they are locally freely-generated by the following formal expressions for $k \in \mathbb{Z}$ fixed and $ i \in \mathbb{N} \cup \{ 0\}$ 
\begin{align}
\Omega^{k;1}_{\proj {1|2}}(U_0) = \mathcal{O}_{\proj {1|2}} (U_0) \cdot  \Big \{  dz d\theta_1^{i-1+k} \delta^{(i)} (d\theta_2), 1 \leftrightarrow 2 \, \Big | d\theta_1^{i+k} \delta^{(i)} (d \theta_2), 1 \leftrightarrow 2  \, \Big \},
\end{align}
so that for example the element $\pi \mathcal{D}[dz| d\theta_1, d\theta_2]$ lifted by $\Theta_1 \equiv \Pi$ from $\mathcal{B}er (\proj {1|2})$ to $\Omega^{0;1}_{\proj {1|2}}$ is given by the choice $k = 0, i = 0$ in the previous expression and corresponds to $dz\delta^{(0)} (d \theta_2)/d\theta_1 ,$ as we have seen.\\
It is possible to jump from the last line of the diagram \eqref{diag} - that of integral forms -, to the first line - that of superforms, comprising also inverse superforms -, by composing the maps $ \Theta_2 \circ \Theta_1 \defeq \Theta (\iota_{\theta_2}) \circ \Theta (\iota_{\theta_1}),$ and we call the morphism resulting from this composition $\Theta_{{max}}$. Formally, this morphism converts both of the delta's appearing in an integral form in two inverse superforms, that is $\delta^{(\ell_1)} (d \theta_1) \delta^{(\ell_2)} (d\theta_2) \longmapsto \frac{1}{d\theta_1^{\ell_1+1} d\theta_{2}^{\ell_2 +1}}$. Notice, that $\Theta_{max}$ is therefore an \emph{even} morphism, and indeed by a sheaf-theoretic approach, looking again at the transition functions by a local computations, one can see that $\Theta_{max}$ acts nothing but the composition of two parity changing functor $\Pi, $ so that one has
\bear
\xymatrix@R=0.1cm{
\Theta_{max} \defeq  \Theta (\iota_{\theta_2}) \circ \Theta (\iota_{\theta_1}) : \Omega^{k;2}_{\proj {1|2}} \ar[rr]^{\qquad \quad \qquad \Pi} && \Omega^{k-1; 1}_{\proj {1|2}} \ar[rr]^{ \Pi}  && \Omega^{k-2;0}_{\proj {1|2}} \\
\qquad \qquad \qquad \qquad \qquad s \ar@{|->}[rr] && \pi s \ar@{|->}[rr] && s,
}
\eear 
where we have denoted by $s$ a generic section. Notice also that $\Theta_{max}$ is injective but certainly not surjective, as indeed a generic $\Omega^{k;0}_{\proj {1|2}}$ is infinitely-generated by formal expression of the kind
\bear
\Omega^{k;0}_{\proj {1|2}}(U_0) = \mathcal{O}_{\proj {1|2}} (U_0) \cdot  \Big \{ d\theta_1^{\kappa_1} d \theta_2^{\kappa_2} \, \Big | dz d\theta_1^{\ell_1} d\theta_2^{\ell_2} \Big \},
\eear
for $k \in \mathbb{Z}$ and $\kappa_1 + \kappa_2 = k$, $\ell_1 + \ell_2 = k-1$. In light of this, $\Theta_{max}$ \emph{injects} all of the sheaves appearing in the complex of integral forms into the extended complex of superforms, that is for $k \leq 1$ we have:
\bear
\Theta_{max} : \Pi Sym^{|k-1|} \Pi \mathcal{T}_{\proj {1|2}} \longhookrightarrow \Omega^{k-2;0}_{\proj {1|2}},
\eear
or pictorially, getting back to \eqref{diag}, we have
\bear \label{diagmax}
\xymatrix@C=1.7cm{
\cdots \ar[r] & \Omega^{-1;0}_{\proj{1|2}} \ar[r] & \Omega^{0;0}_{\proj {1|2}} \ar[r] & \Omega^{1;0}_{\proj {1|2}} \ar[r] & \cdots  \\
\cdots \ar[r] & \Pi Sym^2 \Pi \mathcal{T}_{\proj {1|2}} \ar[r] & \ar@{_{(}->}[ul]_{\Theta_{max}}  \mathcal{T}_{\proj {1|2}} \ar[r] & \ar@{_{(}->}[ul]_{\Theta_{max}} \Pi \mathcal{O}_{\proj {1|2}} \ar[r] & 0. \\
}
\eear
In other words, taking for example a section of the Berezinian sheaf in the formal delta's representation, using the morphisms $\Theta_{1}, \Theta_2$ and $\Theta_{max}$, one moves through the following commutative diagram
\bear
\xymatrix{
& \dfrac{dz}{d\theta_1d\theta_2} \in \Omega^{-1;0}_{\proj {1|2}} \\
\Omega^{0;1}_{\proj {1|2}} \owns \dfrac{dz \delta^{(0)} (d \theta_2)}{d\theta_1}  \ar[ur]^{\Theta_2} \ar@{<->}[rr] & & \dfrac{dz \delta^{(0)} (d \theta_1)}{d\theta_2}  \in \Omega^{0;1}_{\proj {1|2}} \ar[ul]_{\Theta_1}  \\
& \ar[ul]^{\Theta_1} \ar[ur]_{\Theta_2} dz \delta^{(0)} (d\theta_1) \delta^{(0)} (d\theta_2) \in \Omega^{1;2}_{\proj {1|2}}. \ar[uu]_{\raisebox{2.9em}{\scriptsize{$\Theta_{max}$}}}
}
\eear
As already observed, $\Theta_{max} \equiv \Pi \circ \Pi = id$, and as such it maps a section of the Berezinian sheaf $\mathcal{D}[dz |d\theta_1 d\theta_2] \equiv dz \delta^{(0)} (d\theta_1) \delta^{(0)} (d\theta_2) \in \Pi \mathcal{O}_{\proj {1|2}} (U_0)$ to itself, as can be seen by looking at the transformation properties of the formal expression $dz / d\theta_1 d\theta_2.$ The horizontal arrows means that one has a correspondence, $\Theta_1 (\mathcal{D}[dz |d\theta_1 d\theta_2]) = \pi \mathcal{D}[dz |d\theta_1 d\theta_2] = \Theta_2 (\mathcal{D}[dz |d\theta_1 d\theta_2]).$
Both in the sheaves $\Omega^{k;1}_{\proj {1|2}}$ having middle-dimensional picture, and in the extended superform sheaf $\Omega^{k;0}_{\proj {1|2}}$ there are an infinite number of generators that do not come from liftings of generators of the locally-free sheaves $\Omega^{k;2}_{\proj {1|2}}$ via $\Theta_i, $ for $i =1,2$. This is the case, looking for example at $\Omega^{0;1}_{\proj {1|2}}$, of the formal generators $\{ dz d\theta_1^{i-1} \delta^{(i)} (d\theta_2)\}, 1 \leftrightarrow 2 \, | d\theta^{j}_1 \delta^{(j)} (d\theta_2), 1 \leftrightarrow 2 \,\}$ for $i > 0$ and $j \geq 0$ (recall that the case $i=0$ corresponds to the lifting of a generating section of the Berezinian sheaf $\Omega^{1;2}_{\proj {1|2}} = \mathcal{B}er (\proj {1|2})$). These in turn lifts to $\{d\theta_1^{j} d\theta_2^{-j-1}, 1 \leftrightarrow 2\, |\, dz d\theta_1^{i-1} d\theta_2^{-i-1}, 1 \leftrightarrow 2 \}$ for $i > 0$ and $j \geq 0$ in $\Omega^{-1;0}_{\proj {1|2}}$. One possibility that can be put forward in order to explain the insurgence of these infinity amount of somehow \virgolette spurious'' elements is to look at these as \emph{pure gauge}. Indeed, choosing $\Theta (\iota_{1})$ and $\Theta (\iota_2)$ is actually a gauge choice, as taking a contraction $\iota_{i} \defeq \iota_{\partial_{\theta_i}}$ along $\partial_{\theta_i}$ corresponds to pick up a privileged direction. Indeed, it can be seen that just by rotating the element $d\theta_1$ so that $d\theta_1 \rightarrow \alpha_1 d\theta_1 + \alpha_2 d\theta_2$, one gets 
\begin{align}
\frac{1}{d\theta_1} \stackrel{R}{\longmapsto}\dfrac{1}{\alpha_1 d\theta_1 + \alpha _2 d \theta_2} = \frac{1}{\alpha_1d\theta_1 (1 + \frac{\alpha_2}{\alpha_1} \frac{d\theta_2}{d\theta_1}) } = \frac{1}{\alpha_1 d\theta_1} \sum_{k= 0}^\infty (-1)^k \left ( \frac{\alpha_2 }{\alpha_1}\right )^k \left ( \frac{d\theta_2}{d\theta_1} \right )^k,
\end{align}  
and all of the powers of the element $d\theta_2/ d\theta_1$ appears by expanding the rotated elements.\\
In support of this gauge interpretation, notice that this issue is bypassed as one considers $\Theta_{max} = \Theta_1 \circ \Theta_2$ instead, jumping directly to the top line of extended superforms, indeed no choice has been made in this case, as one performs a contraction $\iota_{\partial_{\theta_1}} \circ \iota_{\partial_{\theta_2}}$ along both $\partial_{\theta_1}$ and $\partial_{\theta_2}$, saturating all of the available directions. \\

In this scenario, it is worth pointing out that the relationship between the maps given by $\Theta_i$ and $\Theta_{max}$ in particular, and the map $\eta_0$ introduced in section \ref{LHSPCO}. Actually, working over $\proj {1|2}$, we consider two maps, $\eta_i : \Omega^{k;0} \rightarrow \Omega^{k+1;2}$ for $i=1,2$ and $k\in \mathbb{Z}$ whose action is given in the delta's formalism by 
\bear
\eta_i \defeq \left \{ \begin{array}{ll}
{d\theta_i^k} \longmapsto \delta^{(|k|-1)} (d\theta_i) & k < 0 \\
d\theta^{k}_i \longmapsto 0 & k \geq 0,
\end{array}
\right. 
\eear
which implies that each of the $\eta_i$ has an infinite dimensional kernel: looking at $\Omega^{-1;0}_{\proj {1|2}}$ for example, one sees that all the generators of the form $\{ d\theta_1^j d\theta_2^{-j-1} \}_{i \geq 0}$ gets mapped to zero by $\eta_1,$ and only choosing $i=0$ one gets a non-zero element in $\Omega^{0;1}_{\proj{1|2}}$ by $\eta_2$, namely $\eta_2 (d\theta_2^{-1}) = \delta^{(0)} (d\theta_2)$. Interestingly, just as before for the $\Theta_i$'s, one can consider the composition $\eta_{max} \defeq \eta_1 \circ \eta_2 : \Omega^{k;0}_{\proj {1|2}} \rightarrow \Omega^{k+2;2}_{\proj {1|2}}$, which has been denoted $\eta_0$ in section \ref{LHSPCO} where just one odd dimension was taken into account. Considering a generic sheaf $\Omega^{k;0}_{\proj {1|2}},$ for $k_1 + k_2 = k$ one has that
\bear
\eta_{max} \defeq \left \{ \begin{array}{ll}
d\theta_1^{k_1} d\theta_2^{k_2} \longmapsto \delta^{(|k_1|-1)} (d\theta_i) \delta^{(|k_2|-1)} (d\theta_2)& k_1, k_2 < 0 \\
d\theta^{k_1}_1 d\theta_2^{k_2} \longmapsto 0 &  \mbox{else}.
\end{array}
\right. 
\eear
Keep on looking at the delta's formalism, this implies that $\Theta_{max}$ and $\eta_{max}$ are inverse \emph{up to a kernel}: {i.e.}\ if one the one hand $\Theta_{max} : \Omega^{k+2;2}_{\proj {1|2}} \rightarrow \Omega^{k;0}_{\proj {1|2}}$ is an \emph{injective} and \emph{not surjective} sheaf morphism, conversely, $\eta_{max} : \Omega^{k;0}_{\proj {1|2}} \rightarrow \Omega^{k+2;0}_{\proj {1|2}}$ is a \emph{surjective} and \emph{not injective} sheaf morphism, where once again we recall that $\Omega^{k+2;0}_{\proj {1|2}}$ is a locally-free sheaf of a certain (finite) rank, so that the equation 
\bear
\Theta_{max } \circ \eta_{max} \circ \Theta_{max} = \Theta_{max} 
\eear
holds true as claimed in \cite{Giappo} from string field theory considerations. Notice also that the extended superforms sheaf $\Omega^{k;0}_{\proj {1|2}}$ fits into a short exact sequence as follows 
\bear
\xymatrix{
0 \ar[r] & \ker \eta_{max} \ar[r] & \Omega^{k;0}_{\proj {1|2}} \ar[r] & \mbox{Im}\, \eta_{max} \ar[r] & 0,
}
\eear
or 
\bear
\xymatrix{
0 \ar[r] & \mbox{Im}\, \Theta_{max} \ar[r] & \Omega^{k;0}_{\proj {1|2}} \ar[r] & \mbox{coker}\, \Theta_{max}  \ar[r] & 0,
}
\eear
where we recall that one has a correspondence $\mbox{Im}\, \eta_{max} \cong \Omega^{k+2; 2}_{\proj {1|2}}$ and $\mbox{coker}\, \Theta_{max} \cong \ker \eta_{max}.$\\
Notice also that the free equation of motions for string field theory \cite{Berko, Giappo} can be recovered in this geometric setting by the cohomological equation 
\bear
d ( \mbox{Im}\, \eta_{max}  ) = 0.
\eear
This is well-defined as the operator $\eta_{max}$ maps the infinite dimensional Large Hilbert Space to the Small Hilbert Space, which is a represented sheaf-theoretically by a locally-free sheaf of finite rank.

\section{Serre Duality and Hodge Diamond of a Supermanifold}

\noindent 
It is known, as remembered also early on in this paper, that in general the de Rham cohomology does not yield any information about the supergeometric structure and it coincides with the ordinary de Rham cohomology of the reduced manifold $\manir$ \cite{Deligne, VorGeom}. Things are clearly different for \v{C}ech cohomology instead.
In this context, a very important early result due to Penkov \cite{Pen} states that, if $\mani$ is projective, \emph{i.e.}\ if there exists an embedding morphism $\varphi : \mani \rightarrow \proj {k|l}$, then the \emph{dualizing sheaf} $\omega_\mani$ is given by $\mathcal{B}er (\mani )$ and, just as in the classical commutative case, for any coherent sheaf $\mathcal{F}$ of $\mathcal{O}_\mani$-modules one has the isomorphism of vector superspaces $Ext^i (\mathcal{F}, \mathcal{B}er (\mani)) \cong H^{n-i} (\mani, \mathcal{F})^\ast $ for $i \geq 0$, which is the generalization of \emph{Serre duality} to the context of supergeometry. Notice by the way - in comparison with the ordinary commutative case - that in a supergeometric context it is no longer true in general that $H^n (\mani, \mathcal{B}er (\mani))$ is (even or odd) one-dimensional, isomorphic to $\mathbb{C}$ or $\Pi \mathbb{C}$: just consider for example the case of a split complex supermanifold $\mani$ of dimension $1|1$ over $\proj 1$, whose structure sheaf is given by $\mathcal{O}_{\proj {1}} \oplus \Pi \mathcal{O}_{\proj 1}.$ Computing, one finds that $H^1 (\mani, \mathcal{B}er (\mani)) \cong \mathbb{C}^{1|1}.$ \\
To our limited aims, anyway, we will write Serre duality in the easier form 
\bear \label{sd}
H^i (\mani, \mathcal{F}) \cong \Pi^nH^{n-i} (\mani, \mathcal{B}er (\mani) \otimes \mathcal{F}^\ast)^{\ast},
\eear
for $n$ the even dimension of $\mani$ and $i = 0, \ldots, n$, and where $\Pi^n= \Pi$ if $n$ is odd and $\Pi^n = id$ if $n$ is even. \\
\noindent In the hypothesis we are working with a projective supermanifold $\mani$, if one takes $\mathcal{F}$ to be the sheaf of superforms $Sym^k \Pi \mathcal{T}^\ast_\mani$, Serre duality \eqref{sd} yields the following interesting isomorphism, relating the \v{C}ech cohomology of superforms to the \v{C}ech cohomology of integral forms
\bear
\underbrace{H^{i} (\mani, Sym^k \Pi \mathcal{T}^\ast_\mani)}_{\mbox{\tiny{superforms}}} \cong \underbrace{H^{n-i} (\mani, \mathcal{B}er (\mani) \otimes Sym^k \Pi \mathcal{T}_\mani )^\ast}_{\mbox{\tiny{integral forms}}}
\eear
for $n$ the even dimension of the supermanifold $\mani$, $i = 0, \ldots, n$ and $k \geq 0,$ and where we have forgotten about the parity for simplicity. In other words, the \v{C}ech cohomology of integral forms is fully determined by the \v{C}ech cohomology of superforms and viceversa.\\
It is interesting then to have a look at the supergeometric analog of the \emph{Hodge diamond} of a supermanifold. Working in analogy with the ordinary complex geometric case, one set the \emph{super Hodge numbers} of $\mani$ to be   
\bear
h_s^{p,q} (\mani) = \mbox{dim}_\mathbb{C}\, H^q (\mani, Sym^p \Pi \mathcal{T}^\ast_\mani), 
\eear
where here $\mani$ is again a generic complex supermanifold of dimension $n|m$. There is an obvious - yet striking - difference compared to the ordinary complex geometric case: one finds that in general $h^{p,q}_s (\mani) \neq 0$ for $p > n = \mbox{dim}\, \manir $ since the de Rham complex is not bounded from above. On the other hand, it keeps being true that $h^{p,q}_s (\mani) = 0$ for $q > n$, so that, on the most general ground, the super Hodge diamond, will not really be an actual diamond, but a heavily left-weighted shape instead.
Let us consider for example the case of a supermanifold $\mani$ having even dimension equal to $3$.
\bear
\xymatrix@C=2pt@R=2pt{
\ddots && \ddots && \ddots && \ddots\\	
& h^{7,0}_s & & h^{6,1}_s & & h^{5,2}_s & & h^{4,3}_s \\ 
& & h^{6,0}_s && h^{5,1}_s && h^{4,2}_s && h^{3,3}_s \ar@{-}[dr]\\
& & & h^{5,0}_s & & h^{4,1}_s & & h^{3,2}_s \ar@{-}[ur] && h_s^{2,3} \ar@{-}[dr] \\
& & & & h^{4,0}_s & & h^{3,1}_s \ar@{-}[ur] & & h^{2,2}_s & & h^{1,3}_s \ar@{-}[dr] & & & \\ 
& & & & & h^{3,0}_s \ar@{-}[dr] \ar@{-}[ur] & & h^{2,1}_s & & h^{1,2}_s & & h^{0,3}_s \ar@{-}[dl] & & \\ 
& & & & & & h^{2,0}_s \ar@{-}[dr] & & h^{1,1}_s & & h^{0,2}_s \ar@{-}[dl]& & & \\
& & & & & & & h^{1,0}_s \ar@{-}[dr]& & h_s^{0,1} \ar@{-}[dl] & & & & \\
& & & & & & & & h^{0,0}_s  & & & & &
}
\eear
In the picture above, we have highlighted the region where the ordinary Hodge diamond of the complex manifold $\manir$ is concentrated. Notice anyway that one should refrain to interpret the sum of the super Hodge numbers in each row as the \emph{Betti numbers} $b_k (\mani)$ of the supermanifold $\mani$. The Betti numbers $b_k$ are indeed \emph{topological invariants} and as such they only depend on the topology of the supermanifold - which actually corresponds with the topology of its reduced complex manifold $\manir$ -, whilst super Hodge numbers are finer invariants that heavily depend of the supergeometric structure of the supermanifold $\mani$. That is to say, {homeomorphic} supermanifolds - that yields identical Betti numbers $b_k$ - might possibly give rise to very different super Hodge numbers. It is by the way opinion of the authors that it would be very interesting to generalize Hodge theory and the technology related to the so-called Hodge decomposition theorems to supergeometry: it is actually possible that the sum of super Hodge numbers might acquire some significance in this extended framework. \\
Getting back to superforms, integral forms and Serre duality relating their \v{C}ech cohomologies, it is interesting to restore the \emph{picture number} formalism. Serre duality then reads
$
H^{q} (\mani, \Omega^{p;0}_\mani) \cong \Pi^n H^{n-q} (\mani, \Omega^{n-p;m}_\mani)^\ast,
$
for $p \geq 0$ and $q=0, \ldots, n$, which (up to the parity) is actually what expected by similarity with the ordinary case in complex algebraic geometry in terms of Hodge numbers, but now the difference is that we have to take the picture number into account: $h^{p,q|0}_s (\mani) = h^{n-p, n-q|m}_s (\mani),$ up to parity inversion depending on the dimension. Let us represent this symmetry pictorially, for a certain supermanifolds of dimension $2|m$:
\bear
\xymatrix@C=0.2pt@R=0.1pt{
& \ddots \qquad && \ddots \qquad && \ddots \qquad &&\\	
& h^{4,0|0}_s && h^{3,1|0}_s && h^{2,2|0}_s \\
& & h^{3,0|0}_s & & h^{2,1|0}_s & & h^{1,2|0}_s  \\
& & & h^{2,0|0}_s & & h^{1,1|0}_s  & & h^{0,2|0}_s  \\ 
& & & & h^{1,0|0}_s & & h^{0,1|0}_s \\ 
& & & & & h^{0,0|0}_s \\ 
\\
& & & &h^{2,1|m}_s & & h_s^{1,2|m} & & & & \\
& & &h^{2,0|m}_s & & h^{1,1|m}_s & & h^{0,2|m}_s & & & \\
& & & & h^{1,0|m}_s & & h^{0,1|m}_s & & h^{-1, 2|m}_s \\
& & & & & h^{0,0|m}_s & & h^{-1,1|m}_s & & h^{-2,2|m}_s \\
 & & & & &  \;\; \ddots & & \; \; \ddots & & \; \; \ddots  
}
\eear
Where we have used that in general $h^{0,0|0}_s (\mani) = h^{n,n|m}_s (\mani)$, as to achieve a more symmetric picture. Serre duality is represented as in the ordinary complex geometric context by a rotation by an angle $\pi$ along $h^{0,0|0}_s (\mani)$ relating the upper with the lower figure.
\vspace{.5cm}\\
Let us now consider the case of compact \emph{super Riemann surfaces} $\mathcal{S}\Sigma_{g}$ of a fixed genus $g$ (see \cite{ManinNC}, or also \cite{FioresiKwok} for more details). \\
A compact super Riemann surface $\mathcal{S}\Sigma_g$ of genus $g$ is the data of pair $(\mani^{1|1}, \mathcal{D})$, where $\mani^{1|1}$ is a complex supermanifold such that $\mani_{red}^{1|1} = \Sigma_g,$ where $\Sigma_g$ is a compact Riemann surface, and $\mathcal{D}$ is a locally-direct (and hence locally-free) subsheaf of $\mathcal{T}_{\mani^{1|1}}$ of rank $0|1$ such that $\mathcal{D}^{\otimes 2} \cong \slantone{\mathcal{T}_{\mani^{1|1}}}{\mathcal{D}}$, via $\mathpzc{d}_1 \otimes \mathpzc{d}_2  \longmapsto \{\mathpzc{d}_1 , \mathpzc{d}_2 \} \, \mbox{mod} \,\mathcal{D},$ where $\{ \cdot  , \cdot \}$ is the (super) Lie bracket - notice that $\mathpzc{d}_1$ and $\mathpzc{d}_2$ are sections of $\mathcal{D}$ and as such are \emph{odd} vector fields, so that the super Lie bracket here is actually the anticommutator $\{\mathpzc{d}_1, \mathpzc{d}_2 \} = \mathpzc{d}_1 \mathpzc{d}_2 + \mathpzc{d}_2 \mathpzc{d}_1$.\\
In what follows we will employ an equivalent characterization for super Riemann surfaces, using \emph{theta characteristics} $\Theta_g$ on $\Sigma_g$ (see for example \cite{ACGH}). To this end we first recall that a theta characteristic is an element in $\mbox{Th} (\Sigma_g) \defeq \left \{ \Theta_g \in \mbox{Pic}^{g-1} (\Sigma_g) : \Theta_g^{\otimes 2} \cong \mathcal{K}_{\Sigma_g}  \right \}, $ where $\mathcal{K}_{\Sigma_g}$ is the canonical sheaf of the compact Riemann surface $\Sigma_g :$ in other words a theta characteristic $\Theta_g$ is the data of a pair $(\Theta_g, \varphi_g)$, where $\Theta_g $ is a line bundle on $\Sigma_g$ and $\varphi_g : \Theta_g^{\otimes 2} \rightarrow \mathcal{K}_{\Sigma_g}$ is an isomorphism - this is why a theta characteristic is often denoted as a \virgolette square root'' of the canonical sheaf $\Theta_g = \mathcal{K}^{\otimes 1/2}_{\Sigma_g}.$ \\
In the following we will use that giving a compact super Riemann surface of genus $g$ as above is the same as giving a pair $(\Sigma_g, \Theta_g)$, where $\Sigma_g$ is an ordinary compact Riemann surface of genus $g$ and $\Theta_g$ a theta characteristic on it (see again \cite{ManinNC, FioresiKwok}), so that one can equivalently take this as a definition and indeed we will write $\mathcal{S}\Sigma_g \defeq (\Sigma_g, \Theta_g).$ The structure sheaf of a super Riemann surface $\mathcal{S}\Sigma_g$ is given by $\mathcal{O}_{\mathcal{S}\Sigma_g} = \mathcal{O}_{\Sigma_g} \oplus \Theta_g$, which becomes a sheaf of superalgebras with multiplication law $(f_1, \theta_1 )\cdot (f_2, \theta_2) = (f_1f_2,  f_1 \theta_2 + f_2 \theta_1).$ Clearly, the structure sheaf of a super Riemann surface is a sheaf of $\mathcal{O}_{\Sigma_g}$-algebras: it follows that the sheaf of $1$-forms $\Pi \mathcal{T}_{\mathcal{S}\Sigma_g}^\ast$ is a (locally-free) sheaf of $\mathcal{O}_{\Sigma_g}$-modules as well. More precisely, one finds that
\begin{align}
\Pi \mathcal{T}_{\mathcal{S}\Sigma_g}^\ast & = \Pi \mathcal{T}_{\mathcal{S}\Sigma_g}^\ast \otimes \mathcal{O}_{\mathcal{S}\Sigma_g} \cong \Pi \mathcal{T}_{\mathcal{S}\Sigma_g}^\ast \otimes ( \mathcal{O}_{\Sigma_g} \oplus \Theta_g) 
 \cong \mathcal{K}_{\Sigma_g}^{\otimes 1/2} \oplus \mathcal{K}^{\otimes 3/2}_{\Sigma_g} \oplus \Pi (\mathcal{K}^{\oplus 2}_{\Sigma_g}).
\end{align}
This gives the splitting of the sheaf of $1$-forms of a super Riemann surface in terms of sheaves of $\mathcal{O}_{\Sigma_g}$-modules. In turns, the symmetric powers $Sym^k \Pi \mathcal{T}^\ast_{\mathcal{S}\Sigma_g}$, appearing in the de Rham complex are easily computed from the above expression, and one finds that
\bear \label{symk}
Sym^k \Pi \mathcal{T}^\ast_{\mathcal{S}\Sigma_g} \cong \left ( \mathcal{K}_{\Sigma_g}^{\otimes \frac{k}{2}} \oplus \mathcal{K}_{\Sigma_g}^{\otimes \frac{k+2}{2}}\right ) \oplus \Pi \left ( \mathcal{K}_{\Sigma_g}^{\otimes \frac{k+1}{2}}\right)^{\oplus 2}.  
\eear
Given this decomposition, the cohomology can be computed using \emph{Riemann-Roch theorem} for invertible sheaves over ordinary compact Riemann surfaces, 
\bear
h^{0} (\mathcal{L}_{\Sigma_g}) - h^1(\mathcal{L}_{\Sigma_g} ) = \mbox{deg} (\mathcal{L}_{\Sigma_g}) -g +1,
\eear
where we recall that for $\mathcal{L}_{\Sigma_g} = \mathcal{K}_{\Sigma_g}$, one has $\mbox{deg} (\mathcal{K}_{\Sigma_g}) = 2g-2$, and whenever $\mbox{deg} (\mathcal{L}_{\Sigma_g}) > 2g-2$, one has index of speciality $h^1 (\mathcal{L}_{\Sigma_g}) = 0, $ so that Riemann-Roch simplifies. In particular, with reference to the above equation \eqref{symk}, if one chooses $k > 2$, all of the summands in the decomposition have vanishing index of speciality, so that in particular
\bear
h^{k,1}_s (\mathcal{S}\Sigma_g) = h^1 (Sym^k \Pi \mathcal{T}^\ast_{\mathcal{S}\Sigma_g}) = 0|0 \qquad \qquad k >2,
\eear
and one has 
\bear
h^{k, 0}_s (\mathcal{S}\Sigma_g) = h^0 (Sym^k \Pi \mathcal{T}^\ast_{\mathcal{S}\Sigma_g}) = h^0 (\mathcal{K}^{\otimes \frac{k}{2}}_{\Sigma_g}) + h^0 (\mathcal{K}^{\otimes  \frac{k+2}{2}}_{\Sigma_g}) \, \big |\,  2 h^0 ( \mathcal{K}^{\otimes \frac{k+1}{2}}_{\Sigma_g}) \qquad k >2.
\eear
The global sections can then be computed by Riemann-Roch and in the case $k>2$ one has:
\begin{align}
& h^0 (\mathcal{K}^{\otimes \frac{k}{2}}_{\Sigma_g}) = (k-1)g - k +1, \\
& h^0 (\mathcal{K}^{\otimes  \frac{k+2}{2}}_{\Sigma_g}) = (k+1)g - k - 1,\\
& h^0 ( \mathcal{K}^{\otimes \frac{k+1}{2}}_{\Sigma_g}) = k (g-1),
\end{align}
so that, altogether: 
\bear
h^{k,0}_s (\mathcal{S}\Sigma_g) = h^0 (Sym^k \Pi \mathcal{T}^\ast_{\mathcal{S}\Sigma_g}) = 2k (g-1)\, |\, 2k (g-1)  \qquad \qquad k>2.
\eear
Also, recalling that in general $h^0 (\mathcal{K}_{\Sigma_g}) = g$ and $h^1 (\mathcal{K}_{\Sigma_g} ) = 1$, and that for genus $g\geq2$, $h^{0}(\mathcal{K}_{\Sigma_g}) = 3g-3$ and $h^{0} (\mathcal{K}_{\Sigma_g})= 2g-2$ count the number of the even and odd moduli, one has that that 
\begin{align}
& h^{2, 0}_s (\mathcal{S}\Sigma_{g\geq 2}) =  h^{0} (Sym^2 \Pi \mathcal{T}^\ast_{\mathcal{S}\Sigma_{g\geq2} }) = 4g-3 | 4g-4  \\
& h^{2,1}_s (\mathcal{S}\Sigma_{g\geq 2} ) = h^1 (Sym^2 \Pi \mathcal{T}^\ast_{\mathcal{S}\Sigma_{g\geq 2}}) = 1|0,
\end{align}
restricted to genus $g\geq 2.$\\
The remaining cohomologies depends from the particular theta characteristic chosen (recall there are $2^{2g}$ inequivalent such choices over a compact Riemann surface). Indeed, by Riemann-Roch one sees that $h^0 (\mathcal{K}^{\otimes \frac{1}{2}}_{\Sigma_g}) = h^1 (\mathcal{K}^{\otimes \frac{1}{2}}_{\Sigma_g})$, moreover this dimension is bounded by Clifford theorem on special divisors, yielding that in general 
\bear
h^0 (\mathcal{K}^{\otimes \frac{1}{2}}_{\Sigma_g}) = h^1 (\mathcal{K}^{\otimes \frac{1}{2}}_{\Sigma_g}) \leq \frac{g+1}{2}.
\eear
We define $ h^0 (\mathcal{K}^{\otimes \frac{1}{2}}_{\Sigma_g}) = h^1 (\mathcal{K}^{\otimes \frac{1}{2}}_{\Sigma_g}) \defeq \nu_{\Theta_g}$, with $\nu_{\Theta_g}\leq \frac{g+1}{2}.$ One can see that 
\begin{align}
& h^{1,1}_s (\mathcal{S}\Sigma_{g}) = h^1 (\Pi \mathcal{T}^\ast_{\mathcal{S}\Sigma_g}) = \nu_{\Theta_g} | \, 2, \\
& h^{1,0}_s (\mathcal{S}\Sigma_{g\geq 2}) = h^0 (\Pi \mathcal{T}^\ast_{\mathcal{S}\Sigma_{g\geq 2}}) = \nu_{\Theta_g} + 2g -2 \,|\, 2g,\\
& h^{0,1}_s (\mathcal{S}\Sigma_{g}) = h^1 (\mathcal{O}_{\mathcal{S}\Sigma_{g}}) = g \, | \,\nu_{\Theta_g},	\\
& h^{0,0}_s (\mathcal{S}\Sigma_{g}) = h^0 (\mathcal{O}_{\mathcal{S}\Sigma_{g}}) = 1 \, | \,\nu_{\Theta_g}.
\end{align}
Once this numerology is concluded, the respective dimensions can be inserted in a pictorial representation as above, as to get the Hodge diamond for a super Riemann surface of genus $\geq 2 $. By the way, it is possibly more instructive to represent this Hodge diamond by rotating it $\pi/4$ clockwise, as to get a \emph{tower} better than a diamond:
\bear
\xymatrix@C=0.2pt@R=0.3pt{
& & & & \vdots &&   \vdots  && &&\\	
& & & &  6g-6 | 6g - 6 \ar@{-}[dd]& & 0|0 \ar@{-}[dd] & &   \\ \\
& & & & 4g-3 | 4g-4 \ar@{-}[dd] & & 1|0  \ar@{-}[dd]  & &   \\ \\
& & & & \nu_{\Theta_g} + 2g -2 \,|\, 2g  \ar@{-}[ddddddrr]& & \ar@{-}[ddddddll] \nu_{\Theta_g} | 2  \\  \\ \\
& & & & 1 | \nu_{\Theta_g}  & & g| \nu_{\Theta_g} \\  \\ \\
& & & & \nu_{\Theta_g} | 2  \ar@{-}[dd] & & \nu_{\Theta_g} + 2g -2 \,|\, 2g \ar@{-}[dd] & & & & \\ \\
& & & & 1|0   \ar@{-}[dd] & & 4g-3 | 4g-4 \ar@{-}[dd] & & & \\ \\
& & & &  0|0 & & 6g-6 | 6g - 6 \\
  & & & &  \vdots & &  \vdots  
}
\eear
Let us now look at the upper part of the above tower. In this representation, the difference between the left and right \virgolette wall'' of tower is nothing but the difference between the graded dimensions of the zeroth and the first cohomology group of the sheaf involved, in this case $Sym^k \Pi \mathcal{T}^\ast_{\mathcal{S}\Sigma_g}$ for $k\geq 1$. It can be observed that this difference is a \emph{topological invariant} - just like in the ordinary case one has the Euler characteristic of a certain vector bundle - and it does \emph{not} depend on the particular spin structure chosen, and therefore on $\nu_{\Theta_g}$: as we have seen above, this is a consequence of the  {Riemann-Roch theorem}. In particular, looking at the superforms, one finds
\bear
\chi_s (Sym^k \Pi \mathcal{T}^\ast_{\mathcal{S}\Sigma_g} ) \defeq h^0 (Sym^k \Pi \mathcal{T}^\ast_{\mathcal{S}\Sigma_g}) - h^1 (Sym^k \Pi \mathcal{T}^\ast_{\mathcal{S}\Sigma_g}) = k \left ( 2g-2 \right ) | k \left ( 2g-2 \right ),
\eear
for $ k \geq 1, g \geq 2.$ Using similar argument as above (splitting and Riemann-Roch theorem), one can get the same conclusion for an arbitrary locally-free sheaf over a super Riemann surface.  \\
Notice by the way that in the first case considered above, given by $g=2$, all of the reduced manifolds are \emph{hyperelliptic} Riemann surface and the possibilities for the theta characteristics on them are easily settled. Indeed, one finds in general that $\deg (\mathcal{K}^{\otimes 1/2}_{\Sigma_{g = 2}}) = 1,$ so that by Clifford theorem $h^0(\mathcal{K}^{\otimes 1/2}_{\Sigma_{g = 2}}) \leq \lfloor 3/2 \rfloor.$ This implies that for a choice of an \emph{even} theta characteristic on $\Sigma_{g=2}$ (there are 10 such) one can only have $h^0 (\mathcal{K}^{\otimes 1/2}_{\Sigma_{g=2}}) = 0, $ while for an \emph{odd} theta characteristic (there are 6 such) one can only find $h^0 (\mathcal{K}^{\otimes 1/2}_{\Sigma_{g=2}}) = 1.$ \\
The higher genus case is actually more complicated, as stressed for example in \cite{WittenRS}: in genus $g=3$, for example, $\mathcal{K}^{\otimes 1/2}_{\Sigma_{g=3}}$ has degree 2. A compact Riemann surfaces in genus $g=3$ is either a quartic curve in $\proj 2$ or hyperelliptic. If and only if it is hyperelliptic, a degree 2 line bundle such as $\mathcal{K}_{\Sigma_{g=3}}^{\otimes 1/2}$ admits a 2-dimensional space of global sections. It follows that an even theta characteristic over an hyperelliptic curve of genus $3$ is such that $ \nu_{\Theta_{g=3}} = 2$. If instead $\Sigma_{g=3}$ is a plane quartic curve, then an even theta characteristic on it will be such that $\nu_{\Theta_{g=3}} = 0.$ In the case one chooses an odd theta characteristic, then the only possibility is $\nu_{\Theta_{g=3}} = 1.$

\section*{Acknowledgements} 
We thank L. Castellani, P. Fr\'e, C. Maccaferri and R. Re for fruitful discussions. 
This research is original and has a financial support of the Universit\`a del Piemonte Orientale. 
(Fondi Ricerca Locale).

\bibliographystyle{amsplain}

\end{document}